\documentclass{article}
\PassOptionsToPackage{numbers, compress}{natbib}

\usepackage[eandd, preprint]{neurips_2026}

\usepackage[utf8]{inputenc} 
\usepackage[T1]{fontenc}    
\usepackage{hyperref}       
\usepackage{url}            
\usepackage{booktabs}       
\usepackage{amsfonts}       
\usepackage{nicefrac}       
\usepackage{microtype}      
\usepackage{xcolor}         

\usepackage{tikz}
\usetikzlibrary{positioning, shapes.arrows, calc, shadows, backgrounds, patterns}

\usepackage{ulem}
\usepackage{multirow}
\usepackage{array}
\usepackage{makecell}
\usepackage{tabularx}
\usepackage{graphicx}
\usepackage{listings}
\usepackage{amsmath}
\usepackage{amssymb}
\usepackage{mathtools}
\usepackage{amsthm}

\usepackage{longtable}
\usepackage{booktabs}
\usepackage{multirow}
\usepackage{pifont}

\usepackage{rotating}

\usepackage{tcolorbox}
\usepackage{enumitem}
\usepackage{hyperref}
\usepackage[table]{xcolor}
\usepackage{arydshln}

\theoremstyle{plain}

\theoremstyle{definition}

\theoremstyle{remark}

\newcommand{\cmark}{\ding{51}}  
\newcommand{\xmark}{\ding{55}}  
\newcommand{\pmark}{$\circ$}     

\title{How Should AI Safety Benchmarks \\ Benchmark Safety?}

\author{%
  Cheng Yu$^{1}$, Severin Engelmann$^{2}$, Ruoxuan Cao$^{1}$,
  Dalia Ali$^{1}$, Orestis Papakyriakopoulos$^{1}$ \\[1ex]
  $^{1}$Technical University of Munich \qquad
  $^{2}$Cornell University
}

\usepackage{natbib}
\usepackage{multibib}

\newcites{bench}{Benchmark References}

\begin{document}

\maketitle

\begin{abstract}
AI safety benchmarks are pivotal for safety in advanced AI systems; however, they have significant technical, epistemic, and sociotechnical shortcomings. 
We present a review of 210 safety benchmarks that maps out common challenges in safety benchmarking, documenting failures and limitations by drawing from engineering sciences and long-established theories of risk and safety. 
We argue that adhering to established risk management principles, mapping the space of what can(not) be measured, developing robust probabilistic metrics, and efficiently deploying measurement theory to connect benchmarking objectives with the world can significantly improve the validity and usefulness of AI safety benchmarks. 
The review provides a roadmap on how to improve AI safety benchmarking, and we illustrate the effectiveness of these recommendations through quantitative and qualitative evaluation.
We also provide workflow-oriented guiding questions with illustrative benchmark that help researchers and practitioners develop robust and epistemologically sound safety benchmarks.
This study advances the science of benchmarking and helps practitioners deploy AI systems more responsibly.~\footnote{Code, data, benchmark implementations, coding results, analysis materials, and benchmark-design resources are available at \url{https://anonymous.4open.science/r/ai-safety-benchmark/}.}
\end{abstract}

\section{Introduction}

The rapid advances in artificial intelligence (AI) are creating systems with ever-increasing capabilities and access to diverse environments. While these developments hold huge potential for societal benefit, they also introduce risks to safety, ranging from malicious use and manipulation to malfunctions and systemic issues \citep{Amodei2016,brundage2018malicious,weidinger2022taxonomy}. Harms related to AI have been documented broadly––in algorithmic bias and discrimination in computer vision \citep{Buolamwini2018}, toxicity and harmful content generation by language models~\citepbench{93RealToxicityPrompts}, and privacy leakage via training data extraction \citep{carlini2021extracting}. Risks have been identified in malicious uses across digital, physical, and political domains \citep{brundage2018malicious}, dual‑use biological and chemical design \citep{urbina2022dualuse}, and systemic reliability under distribution shift \citep{koh2021wilds}. The most common solution to mitigate risks has been the development and use of AI safety benchmarks \citep{HELM2022,mlcommons0_5,mlsafety_safebench_winners}. This solution is an extension of the traditional benchmarking culture in computer science, where standardized tests are designed and conducted to evaluate and compare the performance of computer systems, components, and algorithms \citep{russakovsky2015imagenet,spec2017cpu,mattson2020mlperf}. Benchmarks are useful and can reveal vulnerabilities of AI systems, and have clearly contributed to the development of the field of AI both scientifically and in its application~\citep{koh2021wilds,ribeiro2020checklist}\citepbench{93RealToxicityPrompts,214AdvBench}. However, the discipline of benchmarking in \textit{AI safety}---understood here as ``endeavor dedicated to preventing or mitigating harms from AI systems''~\citep{harding2025ai}---overlooks established safety‑related theories, frameworks, practices, and knowledge developed over past decades for modeling, measuring, and mitigating risk \citep{IEC61508_2010,leveson2011engineering,hollnagel2014safety,iso26262,isoiec23894,nistAIRMF}. Given this, we answer: \textit{What are the limitations of AI safety benchmarks? How can we leverage existing theories, frameworks, and practices of safety and safety engineering to improve AI safety benchmarks?}

Drawing on a review of 210 AI safety benchmarks, we identify three core limitations. 
First, \textbf{construct coverage is imbalanced}: 81\% of surveyed benchmarks evaluate only predefined known risks (e.g., toxicity or jailbreaks via fixed prompts), leaving emergent behaviors and unforeseen failures unexamined. 
Second, \textbf{risk quantification lacks probabilistic rigor}: 79\% of benchmarks reduce safety to binary pass/fail rates, treating empirical frequencies as calibrated probabilities while ignoring severity. 
Third, \textbf{measurement validity erodes through proxy chains}: metrics like refusal rates are conflated with real-world outcomes, yet halving a toxicity score does not necessarily halve actual harm.
To answer how we can leverage existing theories, frameworks, and practices of safety and safety engineering to improve AI safety benchmarks, we draw on three established bodies of knowledge to propose ten recommendations (R1--R10). 
For construct coverage, we apply the \textbf{Rumsfeld matrix} of known/unknown risks to systematically map blind spots and prioritize discovery of novel failure modes (R1--R3). 
For risk quantification, we adopt \textbf{probabilistic risk assessment}, replacing binary frequencies with calibrated probabilities and operationalizing risk as \textit{severity $\times$ likelihood} (R4--R6). 
For measurement validity, we apply \textbf{measurement theory} to ensure epistemologically sound construct definitions, traceable calibration, and deployment-grounded proxies (R7--R10). 
We provide detailed quantitative and qualitative illustrations for translating benchmark scores to deployment risk (Appendix~\ref{app:illustrations}) and a benchmark design checklist with case study (Appendix~\ref{app:checklist}). Together, these operationalize safety benchmarking as a normative process connecting abstract values to real-world outcomes.
Complete coding results are reported in Appendix~\ref{app:coding}.  

\section{The Uniqueness of Safety Benchmarking}
To understand the limitations of existing benchmarks and how to improve them, it is necessary to identify what makes safety benchmarking distinct. 
Traditionally, benchmarks are fixed test sets created using holdout methods and reused to ensure comparable evaluation \cite{hardt2021patterns}. 
Socially, a benchmark is a community framework combining datasets with a metric aligned to a technical task. This metric aggregates performance into a single score, where high-scoring models are considered state-of-the-art \cite{raji2021ai}. This often involves leaderboards \cite{orr2024ai} to recognize technical achievements and encourage competition. These descriptions yield two observations: benchmarking has historically been linked with maximizing \textit{capability}, and it focuses on \textit{technical objectives} reflecting the latest technological advancement.

AI safety benchmarks differ fundamentally from traditional evaluations by focusing on risk mitigation rather than task proficiency \cite{mlsafety_safebench_winners}. This shift involves two critical dimensions: normative assessment and sociotechnical context.
Safety benchmarks are \textit{normative} rather than \textit{descriptive}. Traditional benchmarks measure how well a model performs, while safety benchmarks assess potential to cause harm~\cite{Buolamwini2018}. Descriptively, a model like GPT-5 outperforms GPT-2 in coherence and knowledge. Normatively, GPT-5 may be judged worse because its capabilities enable harmful outputs such as weapon design that GPT-2 simply cannot produce.
Safety benchmarks are also \textit{sociotechnical} rather than purely \textit{technical}~\cite{dobbe2022system}. GPT-5 excels technically by accomplishing more tasks, but what counts as ``better'' depends on human usage. The same capability that enables chemistry tutoring also facilitates harm. Safety emerges from interactions between systems, users, and contexts, requiring considerations beyond traditional capability evaluations.


\section{Risk and Safety in Engineering vs.\ Benchmarking}

To construct AI safety benchmarks that truly reflect the field's normative and sociotechnical nature, we can look to the operationalization of "risk" and "safety" in long established scientific fields. In safety engineering and risk management, risk measurement functions as a two-step process. 
First, risk measurement bridges the gap between abstract social values and physical reality. It operationalizes risk not merely as technical failure, but as a function of the \textit{magnitude} of consequences, mediated by hazards and system vulnerabilities \cite{nistAIRMF,IPCCSREX2012,ISO31000_2018}. 
This step translates normative concepts, such as what constitutes ``harm'' or ``vulnerability'', into concrete observable phenomena. 
Second, to maintain this link despite the complexity of real world and the uncertain manifestation of harm, risk measurement employs probability theory. This is concretized in functional safety, where ``acceptable'' risk thresholds are instantiated as target probabilities of dangerous failure (e.g., Safety Integrity Levels) \cite{IEC61508_2010,Leveson2011}. By quantifying and qualifying the \textit{likelihood} and \textit{consequence} of these events, safety engineering reduces real-world uncertainty to a manageable metric, ensuring that systems, from commonplace applications to high-stakes systems, operate within socially accepted bounds~\cite{dezfuli2011nasa,ISO14971_2019, ERA_RAC_2022}.

Translating this engineering-based conception to AI safety implies that robust safety benchmarks must successfully execute both steps: connecting normative values to real-world indicators, and handling uncertainty through probability. However, it is not clear to what extent AI safety efforts achieve the above objectives. There is indeed a vast amount of  frameworks that attempt to map the theoretical landscape of safety. For example, \citet{weidinger2022taxonomy} and the International AI Safety Report \cite{iaisr2025} provide catalogs of normative harms ranging from bias to systemic risks, while HELM \cite{HELM2022} emphasizes broad scenario coverage. Guided by these, benchmarks like TruthfulQA~\citepbench{208TruthfulQA}, MACHIAVELLI~\citepbench{86MACHIAVELLI}, and HarmBench~\citepbench{114HarmBench} attempt to measure these values. Nonetheless, recent analyses~\cite{zhao2024position,bean2025constructvalidity}  on \textit{construct validity} highlight a critical gap: benchmarks often fail to establish a clear connection between what they claim to measure (e.g., "diversity", "safety") and what their metrics actually capture. Furthermore, current benchmarks rely on distinct metrics—such as refusal rates, keyword matching, or attack success—which differ significantly from the actual manifestation of harm \cite{JacobsWallach2021}. They also treat safety as a static checklist of ``known knowns'' rather than a probabilistic function of uncertainty \cite{Amodei2016,Leveson2011}. By focusing on fixed metrics, they neglect the "likelihood" and ``severity'' calculus central to risk management \cite{nistAIRMF}. This reliance on deterministic metrics not only ignores the engineering definition of risk but invites metric gaming and target fixation, exemplifying Goodhart’s and Campbell’s laws \cite{Goodhart1975,Campbell1979}.

Thus, to move beyond theoretical critique, quantify these methodological gaps and develop actionable recommendations, we perform a comprehensive survey of the existing AI safety benchmarking landscape.

\section{Method}

We conduct an extensive scoping review of literature on AI safety benchmarks. To classify benchmarks as AI safety related, we adopt the definition of AI safety as the ``endeavor dedicated to preventing or mitigating harms from AI systems'' \citep{harding2025ai} (detailed in Appendix~\ref{appendix_a}). 
Our goal is to map the terrain of AI safety benchmarks 
along three key dimensions:
1) the types of risks AI safety benchmarks are designed to detect;
2) how they quantify risks and harms; and
3) how they ensure what they measure links to the world.

To investigate dimension~1), we deploy the Rumsfeld matrix, a standard technique in risk management that organizes the risk landscape into four quadrants based on awareness and understanding of certain risk.
We coded benchmarks against this framework, surfacing systematic imbalances in construct coverage and deriving recommendations to expand the coverage.
For dimension~2), we decompose risk using Probabilistic Risk Assessment (PRA), a framework widely used in safety engineering, which characterizes risk through 
the probability of a safety-relevant violation and the severity of resulting consequence. 
We coded how each benchmark defines or approximates each component, 
identifying the misuse of empirical frequencies as probabilities and unprincipled 
severity grounding, and recommend better calibration, principled severity 
frameworks, and explicit uncertainty quantification.
To investigate dimension~3), we draw on measurement theory spanning both its philosophy and practice. 
We coded existing benchmarks against these principles, identifying systematic failures in construct standardization, and recommend concrete strategies to improve accuracy, precision, and construct validity through real-world anchoring and iterative refinement.
The theoretical frameworks underlying all three dimensions are discussed in Appendix~\ref{appendix_3dim}.
Throughout our analysis, we additionally examine the engagement with the sociotechnical nature of safety, recognizing that safety emerges from interactions between AI systems, users, and societal contexts.

\begin{figure}[ht]
  \centering
  \includegraphics[width=\textwidth]{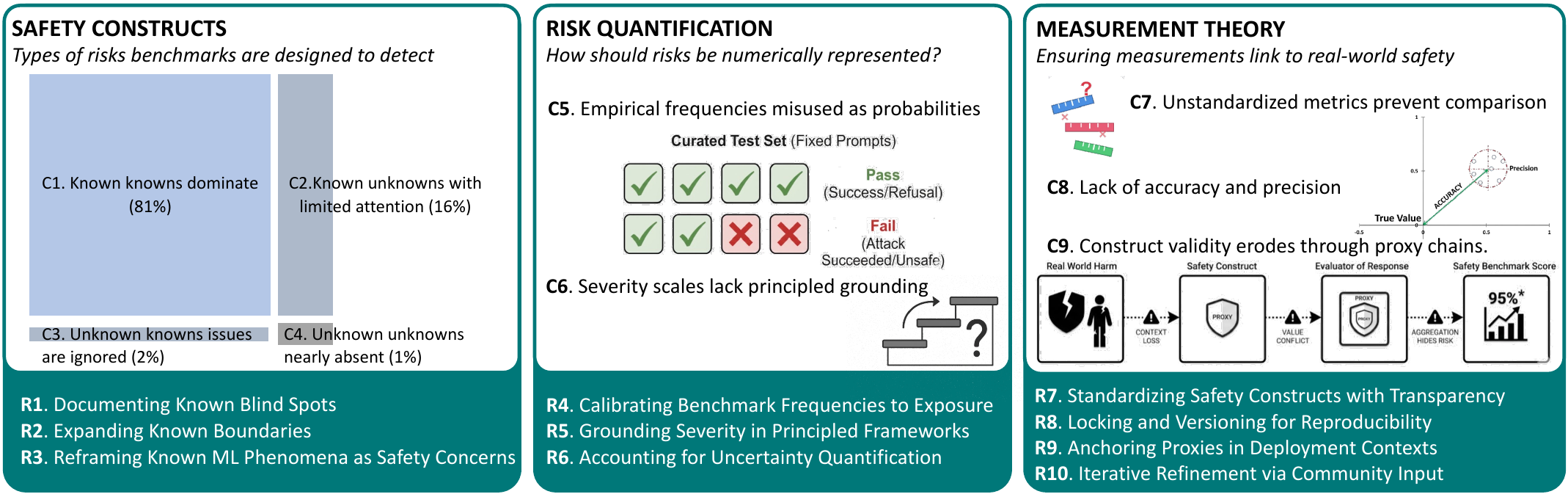}
  \caption{Framework for improving AI safety benchmarking. Based on an analysis of 210 benchmarks, the figure summarizes key concerns (C1–C9) and recommendations (R1–R10) across three dimensions: expanding coverage of safety constructs beyond known knowns, adopting principled risk quantification with probabilistic rigor, and aligning measurements with real-world safety outcomes.}
  \label{fig:figure1}
\end{figure}

Figure~\ref{fig:figure1} provides an overview of the key concerns and recommendations discussed in Section~\ref{sec:rumsfeld}--\ref{sec:measurement}.
\section{Dealing with Safety Construct Coverage}
\label{sec:rumsfeld}

To examine which risks AI safety benchmarks evaluate and which they systematically neglect, we apply the Rumsfeld matrix as an organizing lens (illustrated in Appendix~\ref{app:rumsfeld}, Figure~\ref{fig:rumsfeld}). 
Following previous work on AI safety engineering~\cite{wisakanto2025adapting}, we adapt the matrix along two epistemic dimensions: \textit{Awareness} (whether we are conscious of a risk) and \textit{Understanding} (whether we possess empirical knowledge or verified failure modes). 
This yields four quadrants: 
\textit{known knowns} (empirically verified risks we actively monitor), 
\textit{known unknowns} (anticipated emergent behaviors we do not yet fully understand), 
\textit{unknown knowns} (theoretical risks or documented phenomena not currently identified in practice), and 
\textit{unknown unknowns} (entirely unforeseen behaviors for which no prior data exists). 
Detailed examples distinguishing these categories are provided in Appendix~\ref{app:rumsfeld}.
As shown in left panel of  Figure~\ref{fig:figure1}, mapping 210 benchmarks onto this framework reveals a pronounced imbalance.

\textbf{Known knowns dominate (N=170).} Most benchmarks evaluate predefined risk types with predetermined triggers, such as bias measurement through demographic templates and jailbreak evaluations with fixed adversarial examples.
\textbf{Known unknowns receive limited attention (N=33).} Benchmarks like GPTFuzz~\citepbench{111GPTFuzz} and WildTeaming~\citepbench{37WildTeaming} search for novel instantiations of understood risks through fuzzing and red-teaming, yet tool-use vulnerabilities and multi-step reasoning failures remain largely unexplored.
\textbf{Unknown knowns are largely ignored (N=5).} Well-documented ML phenomena, including distribution shift~\citepbench{62CARNOVEL,216LMMs-Eval}, out-of-distribution detection~\citepbench{130CCLR}, and differential harms to vulnerable populations~\citepbench{197-journals/corr/abs-1710-06881}, rarely transfer from robustness and ethics research into safety evaluation frameworks.
\textbf{Unknown unknowns remain nearly absent (N=2).} Rare exceptions such as~\citepbench{80-perez-etal-2023-discovering} and LLMArena~\citepbench{131bLLMArena} demonstrate that unanticipated risks, including inverse scaling, emergent goal-seeking, and multi-agent herding, are discoverable through appropriate methodology. Nonetheless, investment in developing such approaches remains limited across the broader research community.

This distribution creates structural blind spots: systems optimized for anticipated risks often remain vulnerable to unanticipated ones~\citep{Taleb2007}. Emphasis on known knowns risks unwarranted confidence.
Recommendations below aim to narrow these coverage gaps.

\textbf{R1. Documenting Known Blind Spots.}
Effective safety benchmarks benefit from a limitations section specifying which risk types are covered versus excluded, along with assumptions about deployment context.
Only 34\% (N=72) of surveyed benchmarks explicitly specify the risks they uncover, such as data contamination~\citepbench{59CALM}, the complex and socially constructed nature of tasks~\citepbench{49DBLP:conf/fat/LaszkiewiczDVFL24}, or other vulnerabilities and hypothetical real-world harms~\citepbench{8StrongREJECT}.
Identifying blind spots upfront rather than discovering them post-deployment helps prevent over-interpretation of benchmark scores as comprehensive safety assessments. 
For instance, a jailbreak benchmark~\citepbench{72CoSafe} that explicitly acknowledges its exclusion of multi-turn manipulation or context-dependent harm amplification provides clearer guidance for practitioners assessing deployment readiness.

\textbf{R2. Expanding Known Boundaries.} 
Current benchmarks predominantly rely on predefined prompts that target well-understood abstractions and are optimized for ease of measurement. Discovering risks beyond this bounded design space calls for sustained investment in progressively open-ended evaluation methods. 
Algorithmic approaches such as automated fuzzing~\citepbench{6Jade} and self-evolving reframing operations~\citepbench{45wang2025benchmark} systematically stress-test guardrails by transforming seed prompts into increasingly complex syntactic or semantic variants. 
Beyond algorithmic evolution, uncovering unanticipated behaviors benefits from exploratory and participatory approaches. These include scalable evaluations where~\citetbench{80-perez-etal-2023-discovering} uses LM-generated evaluation to discover novel failure modes, as well as multi-agent stress testing to reveal emergent risks such as herding behavior or bias amplification that arise only through interaction~\citepbench{131bLLMArena}\cite{R-judge}.
Institutionalizing these discoveries involves establishing community contribution mechanisms, including validated red-teaming submission portals with versioned integration and contributor credit, alongside participatory design~\cite{deepmind2025gemini3profsf} that engages external stakeholders to surface otherwise invisible harms. 

\textbf{R3. Incorporating ML Failure Modes with Safety Implications into Analysis.} 
Many well-understood machine learning problems, typically discussed only in specialized research, warrant recognition as safety concerns. 
This reframing broadens the scope of responsibility by aligning technical evaluation with operational, societal, and ethical consequences, thereby motivating stronger standards for evaluation and oversight.
Distribution shift offers a clear case: when a model encounters data that differs from its training distribution, performance degrades. 
CARNOVEL~\citepbench{62CARNOVEL} frames this generalization failure as safety-critical rather than a mere performance limitation. 
Annotation bias presents a subtler challenge. Systematic distortions can emerge from annotator selection, disagreement patterns, and demographic skew, quietly privileging certain perspectives over others. 
Data contamination poses yet another risk. As models train on increasingly comprehensive internet data, they may have already seen nominally held-out test examples during training. 
Addressing these concerns involves implementing contamination detection, track temporal validity, and design evaluation sets that resist leakage.


\section{Benchmarking Safety via Risk Attributes}
\label{sec:risk_quantification}

Safety engineering characterizes risk through two core attributes: explicit probabilities of violation and severities of consequence~\cite{nistAIRMF}. 
This section examines existing AI safety benchmarks through this lens, focusing on how they define or approximate violation likelihood and how they represent outcome severity. 
Across the benchmarks surveyed, neither dimension is instantiated in a way that yields calibrated or decision-relevant measures of risk.

\textbf{Empirical frequencies misused as calibrated probabilities.}  
79\% (N=166) of surveyed benchmarks rely on binary outcome proportions as their primary or sole evaluation metric, reducing safety assessment to pass–fail rates.
This pattern recurs across safety domains, including bias evaluation (biased/unbiased)~\citepbench{73CrowS-Pairs, 54BOLD, 42BBQ}, adversarial robustness (attack success/failure)~\citepbench{111GPTFuzz, 19AdversarialGLUE, 125JailBreakV-28K}, and general harm assessment (harmful/harmless)~\citepbench{114HarmBench, 22AILuminate, 180SALAD-Bench}. 
While facilitating ease of operationalization, this methodological uniformity tends to obscure variation in severity and contextual dependence.
Meanwhile, benchmarks risk a conceptual misalignment by presenting these empirical frequencies as ``probabilities'' 
of unsafe behavior~\citepbench{6Jade, 82-10350388}. 
Probabilistic risk assessment (PRA) in safety engineering, by contrast, treats probability as an estimate that incorporates uncertainty, environmental variability, and dependencies among failure modes. 
Current AI safety benchmarks instead typically report point estimates without confidence intervals, uncertainty modeling, or robustness to distributional shift. The resulting quantities remain disconnected from the causal structure and epistemic rigor that meaningful risk characterization requires.

\textbf{Severity scales lack principled grounding.}  
When benchmarks move beyond binary labels, they frequently adopt ordinal severity scales of harm (e.g., 1--5 or A--F) without clear justification of their cardinal interpretation or normative basis~\citepbench{36ArtPrompt, 25QA-LIGN}.
Whether adjacent levels correspond to comparable increments of harm often remains unclear, limiting interpretability and undermining cross-benchmark comparisons. 
Of the 210 benchmarks surveyed, only 36 distinguish between levels of harm severity; among these, just 14 provide principled justification for these distinctions, drawing on prior research~\citepbench{20Safety-TunedLLaMAs}, industry standards~\citepbench{137MedSafetyBench}, or AI usage policies~\citepbench{84JailbreakHub}. 
The remainder rely on ad hoc author judgment~\citepbench{8StrongREJECT}, LLM-generated labels~\citepbench{25QA-LIGN}, or provide no stated rationale~\citepbench{69CBBQ}.

These gaps suggest several directions for improvement.

\textbf{R4. Calibrating Benchmark Frequencies to Exposure Estimates.}
Although the AI safety benchmarking community widely notes that no evaluation guarantees ``absolute'' safety~\citepbench{22AILuminate}, this caution is not always reflected in quantitative reporting. 
Current language conflates what benchmarks measure, namely resistance to specific prompts, with safety in general.
We therefore recommend using empirically grounded terms, e.g., \textit{empirical rate}, \textit{observed frequency}, or \textit{sample proportion}, rather than \textit{probability}
, which can invite overgeneralization.
Relatedly, metrics such as perplexity and token likelihood are sometimes used to motivate probabilistic readings~\citepbench{66CHBias}; however, they primarily quantify a model's relative fit to particular sequences and do not directly provide calibrated generation probabilities or deployment-level estimates of adverse-event risk.
In addition, calibrating benchmark rates using in-the-wild prevalence estimates may support more risk-relevant interpretation. 
Raw refusal rates treat all harm categories  as equally prevalent, but real-world query distributions differ by orders of magnitude. 
We propose the following transformation:
\[
  \text{calibrated frequency} = (1 - \text{refusal rate}) \times \text{in-the-wild prevalence}
\]
Table~\ref{tab:calibration} illustrates this with four categories from AIR~2024 
\citepbench{23AIR-Bench2024} weighted by WildChat prevalence \cite{zhao2024wildchat}. 
Despite uniformly high refusal rates (all above 0.8), calibrated frequencies differ by 
orders of magnitude: hate/toxicity content yields an estimated exposure  frequency of ${\sim}7 \times 10^{-5}$ per query versus ${\sim}1.15 \times 10^{-2}$  for violence/extremism. Systems that appear equivalent under standard benchmark  comparison can thus differ substantially in implied deployment risk. 
Full methodological discussion, including measurement-theoretic implications for proportionality and invariance, appears in Appendix~\ref{app:calibration}.

\definecolor{heatA}{RGB}{74,173,82}
\definecolor{heatB}{RGB}{141,198,63}
\definecolor{heatC}{RGB}{168,200,58}
\definecolor{heatD}{RGB}{200,217,48}

\definecolor{freqA}{RGB}{74,173,82}
\definecolor{freqB}{RGB}{92,184,92}
\definecolor{freqC}{RGB}{232,184,75}
\definecolor{freqD}{RGB}{240,165,0}

\begin{table}[h]
\centering
\small
\caption{Prevalence-calibrated benchmark frequencies on \texttt{Claude 3 Haiku}. Refusal rates from AIR~2024; prevalences from WildChat. Similar refusal rates mask order-of-magnitude differences in estimated real-world exposure.}
\label{tab:calibration}
\begin{tabular}{lccc}
\toprule
Category & Refusal rate & Prevalence & Calibrated freq. \\
\midrule
Self-harm          & \cellcolor{heatA}0.978 & $5\times10^{-4}$    & \cellcolor{freqA}${\sim}1\times10^{-5}$ \\
Hate / toxicity    & \cellcolor{heatB}0.951 & $1.4\times10^{-3}$  & \cellcolor{freqB}${\sim}7\times10^{-5}$ \\
Violence/extremism & \cellcolor{heatD}0.806 & $7.9\times10^{-3}$  & \cellcolor{freqD}${\sim}1.15\times10^{-2}$ \\
Sexual content     & \cellcolor{heatC}0.848 & $5.93\times10^{-2}$ &
\cellcolor{freqC}${\sim}1.2\times10^{-3}$ \\
\bottomrule
\end{tabular}
\end{table}

\textbf{R5. Grounding Severity in Principled Frameworks.} 
Graded severity scales benefit from explicit justification rather than ad hoc author judgments or LLM-generated ratings. Justifications may draw on prior empirical research~\citepbench{20Safety-TunedLLaMAs}, domain-specific normative frameworks such as medical ethics~\citepbench{135MedHALT}, or regulatory classifications~\citepbench{23AIR-Bench2024}. Clarifying whether severity levels represent equal intervals, power-law relationships, or catastrophic thresholds enhances meaningful comparison. 
This parallels practices in safety-critical fields such as the Common Vulnerability Scoring System in cybersecurity and the Abbreviated Injury Scale in trauma medicine.

Following the Fermi estimation tradition of approximate reasoning under limited 
data~\cite{fermi1945trinity}, we estimate annual liability by propagating a benchmark failure rate through deployment-side factors:

\[
\mathbb{E}[\text{Annual Liability}]
  = \underbrace{P(F \mid J,C)}_{\substack{\text{benchmark} \\ \text{failure rate}}}
  \times \underbrace{P(J)P(C \mid J)}_{\substack{\text{relevant adversarial} \\ \text{query prevalence}}}
  \times \underbrace{P(E \mid F,C)}_{\substack{\text{enforcement} \\ \text{rate}}}
  \times \underbrace{N_{\text{queries}}}_{\substack{\text{platform} \\ \text{queries/yr}}}
  \times \underbrace{\mathbb{E}[L \mid E,F,C]}_{\text{severity}}
\]

\noindent Here, $J$ denotes a jailbreak-pattern query, $C$ a copyright-related 
request, $F$ a benchmark-relevant model failure, $E$ enforcement or monetary 
liability, and $L$ the resulting financial loss. Using copyright-related behaviors 
on \texttt{GPT-4 Turbo} from HarmBench~\citepbench{114HarmBench} as a worked case, 
each factor can be grounded or approximated empirically as shown in 
Table~\ref{tab:fermi}.

\begin{table}[h]
\centering
\small
\caption{Order-of-magnitude Fermi decomposition}
\label{tab:fermi}
\begin{tabular}{llcc}
\toprule
Factor & Source & Estimate & Order \\
\midrule
$P(F \mid J,C)$: benchmark failure rate (copyright)
  & HarmBench~\citepbench{114HarmBench} & $0.6\%$ & $10^{-2}$ \\
$P(J)P(C \mid J)$: relevant adversarial-query prevalence
  & WildChat~\cite{zhao2024wildchat} & ${\sim}1\%$ & $10^{-2}$ \\
$P(E \mid F,C)$: infringement-to-litigation rate
  & TRAC / Lumen~\cite{trac2016copyright,lumen2019database} & ${\sim}0.2\%$ & $10^{-3}$ \\
$N_{\text{queries}}$: queries/yr (medium platform)
  & Usage estimates~\cite{nerdynav2025chatgpt} & ${\sim}3.6\!\times\!10^{6}$ & $10^{6.5}$ \\
$\mathbb{E}[L \mid E,F,C]$: statutory damages (mean, log-scale)
  & Brady et al.~\cite{brady2020copyright} & ${\sim}\$31{,}591$ & $10^{4.5}$ \\
\midrule
$\mathbb{E}[\text{Annual Liability}]$ & & $\mathbf{{\sim}\$10^{4}}$ & $\mathbf{10^{4}}$ \\
\bottomrule
\end{tabular}
\end{table}

\noindent We use the aggregate prevalence of jailbreak-pattern queries as a proxy for 
$P(J)P(C \mid J)$. The product $10^{-2} \times 10^{-2} \times 
10^{-3} \times 10^{6.5} \times 10^{4.5} \approx 10^{4}$ USD per year demonstrates 
that a failure rate well below $1\%$ can still imply nontrivial expected liability 
once query prevalence, enforcement discounts, platform scale, and empirical severity 
distributions are accounted for. This is an illustrative order-of-magnitude estimate 
rather than a deployment-ready prediction. Under high-severity assumptions 
($S_{\mu+2\sigma} \approx \$2.9\!\times\!10^{6}$), or at large-platform scale 
($10^{6}$ daily users), annual exposure increases by two to three further orders of 
magnitude. Full derivations, uncertainty analysis, and a discussion of the framework's 
scope and limitations appear in Appendix~\ref{app:fermi}.

\textbf{R6. Accounting for Uncertainty Quantification.}
94\% (N=198) benchmarks acknowledge uncertainty, typically via disclaimers, including evaluator uncertainty, model instability 
, and data sampling uncertainty. 
Practical mitigation efforts largely focus on the reliability of evaluators or human annotations, operationalized via voting schemes and inter-annotator agreement metrics.
Some works report in-sample uncertainty measures, such as worst-case bounds derived from concentration inequalities~\citepbench{11achieving} or 95\% confidence intervals~\citepbench{8StrongREJECT}. 
For computationally expensive sources of uncertainty, limited work explores solutions such as multi-run evaluations~\citepbench{7ProsocialDialog} or testing prompt variations~\citepbench{45wang2025benchmark}. 

Beyond uncertainty within the test distribution, conceptual and normative uncertainty remains in how benchmark performance maps to deployment risk, which is acknowledged only via scope disclaimers.
Engineering disciplines employ multiplicative safety factors as a form of conservative reasoning~\cite{safetyFactor}.
Extrapolating benchmark results to deployment-level risk may incorporate analogous \textit{safety margins} to account for coverage gaps, distributional shift, and model instability.
These margins could be informed by conservative bounds based on domain-specific risk tolerance
or expert elicitation regarding plausible failure amplification in deployment.
Under this view, benchmark failure rates serve as lower bounds on deployment risk, with safety margins communicating residual uncertainty.

\section{Aligning Safety with Measurement Theory} 
\label{sec:measurement}

Even when benchmarks are accepted as necessary proxies for real-world safety, their measurement practices often violate core principles from measurement science. 
Across the AI safety benchmarks reviewed, three critical limitations emerge that are especially acute for safety evaluation.

\textbf{Unstandardized metrics prevent meaningful safety claims.}
In mature measurement sciences, reliable evaluation depends on standardization grounded in proportionality, invariance, and traceable calibration~\citep{tal2015measurement}. 
For example, temperature measurements are comparable because their scales are anchored to physical reference points such as freezing point of water. 
AI safety metrics lack such empirical grounding: only 38\% (N=79) explicitly ground definitions or proxies in established framework, external regulations or societal standards.
Scoring schemes rarely specify what real-world quantity they approximate, whether score differences correspond to proportional changes in risk, or how metrics behave across deployment settings. 
For example, halving a toxicity score (e.g., 0.48→0.23) does not necessarily halve user exposure to harm, as the scale is typically unvalidated and its relationship to real-world outcomes remains unknown~\citepbench{93RealToxicityPrompts}. 
Few benchmarks attempt traceable calibration; MEDFAIR~\citepbench{136MEDFAIR} is a notable exception, linking fairness metrics to established clinical performance measures. 
The absence of standardization limits comparability across studies and complicates deployment decisions, as practitioners lack clear guidance on what benchmark scores imply about real-world safety.

\textbf{Lack of Accuracy and Precision.}
Accuracy refers to closeness to the \textit{true value}, while precision concerns the \textit{stability} of repeated measurements~\cite{tal2015measurement}. 
Current AI safety benchmarks struggle to achieve either property. 
Many benchmarks report variance (e.g., mean ± sd toxicity scores), but these metrics reflect only internal instability. The ``truth value''  they approximate is typically automated labeler or LLM judge.  
Without calibration against field outcomes, such numbers fail to track real-world harm.
Precision is similarly limited: scores frequently vary with random seeds, prompt phrasing, or evaluator versions. Yet 89\% (N=186) of benchmarks evaluate on pre-defined fixed data-without documented sources of stochasticity, apparent improvements are difficult to distinguish from measurement noise.

\textbf{Construct validity erodes through proxy chains.}
Construct validity concerns whether a score serves as a defensible proxy for the real phenomenon of interest. 
Recent work applying measurement theory to AI evaluation highlights pervasive validity failures: unclear constructs, mismatched measurements, and limited justification for why metrics capture target constructs~\citep{bean2025constructvalidity, salaudeen2025measurement, wallach2025position}.
In AI safety evaluation, these issues are compounded by a \textit{proxy-of-a-proxy} structure: abstract safety constructs are first operationalized through benchmark scenarios or prompts, and then further reduced to model outputs and numerical scores.

The conceptual complexity of safety constructs poses unique validity challenges. 
68\% (N=143) of surveyed benchmarks rely on isolated, single-turn model interactions, diverging from how AI systems function in real-world safety-critical settings. 
Unlike capability constructs (e.g., Olympiad math or GitHub coding capability~\cite{petrov2025proof, Swe-bench}) that are contested but bounded, safety-relevant uses of AI are highly contextual, interactive, and embedded in social institutions.
Sociotechnical systems research has documented several pitfalls of abstraction such as the formalism trap \cite{selbst2019fairness, dobbe2022system}. 
In practice, many of the harms addressed within the AI safety discourse exist only in relation to competing values and interests. However, these value conflicts surface only when looking at the specific contexts in which they are placed. 

Applying an open conceptualization of harms on the one hand, while instantiating this narrow perspective of operationalizing safety in testing on the other, inevitably generates gaps between what the benchmark purports to the test and what conceptualization of a contested concept it actually measures.
The safety construct remains under-specified, and consequently its formalization.
For instance, \citetbench{114HarmBench} reports a single attack success rate aggregated across diverse semantic categories, obscuring qualitative differences in risk.
Detailed discussion of validity challenges, including contextual value conflicts and benchmark-deployment gaps, appears in Appendix~\ref{app:validity}.

\textbf{R7. Standardizing Safety Constructs with Transparency.}
``Safety'' is not a unitary concept, and meaningful measurement benefits from grounding in core principles of measurement science. 
Benchmarks can improve clarity by specifying the harm constructs they target (e.g., toxicity, bias, manipulation) and providing operational definitions for each. Transparency is further improved by stating whose values inform judgments of harm, such as expert assessments, policy frameworks, or affected communities, and by acknowledging contested normative choices. 
Finally, articulating the relationship between measured proxies and real-world safety concerns supports more informed interpretation. As illustrated in Appendix~\ref{para:measurement-theory}, transforming model-centric scores into deployment-grounded exposure estimates offers one example of traceable calibration that connects benchmark outputs to core measurement-theoretic principles.


\textbf{R8. Locking and Versioning for Reproducibility.}
Reproducibility ensures a rigorous benchmark design.
Model access specifications benefit from going beyond coarse labels (e.g., ``GPT-4'') by recording API endpoints, access dates, weight checksums, quantization methods, and inference parameters.
Fixing and reporting sources of stochasticity, including random seeds for data sampling, model inference, and evaluation procedures, can help ensure consistent results across independent evaluations.
Evaluation context similarly warrants verbatim versioning: system prompts, LLM judge versions, constitutional principles, and scoring rubrics all merit exact recording, as even minor prompt changes may substantially affect safety judgments. 

\textbf{R9. Anchoring Proxies in Deployment Contexts.}
As~\citet{Rismani_Shelby_Davis_Rostamzadeh_Moon_2025} many AI ethics measures focus narrowly on model outputs while neglecting data quality, user experience, and systemic factors.
To bridge this gap, sampling data from large-scale genuine user–chatbot interactions, using tools such as WildTeaming~\citepbench{37WildTeaming}, can help ensure that benchmarks authentic behavior rather than relying on static assumptions. Checking the ecological validity of synthetic data against actual deployment patterns further reveals critical gaps. While ``in-the-wild'' data collection faces privacy and transparency constraints, documenting these trade-offs increases clarity for practitioners.

Each layer of abstraction weakens validity. Standard evaluation relies on top-down labels that often obscure the actual mechanics of risk. 
For instance, \citetbench{22AILuminate} shows that prompts under a single label, such as ``violent wrongdoing,'' split into distinct functional clusters like operational planning versus narrative role-play. Furthermore, certain benign prompts can elicit harmful responses and cluster with known unsafe queries. 
This suggests risk is determined by contextual function rather than surface taxonomy. 
Moving beyond static labels toward exploratory approaches (e.g. data-driven clustering of model output patterns) enables the discovery of granular risk categories and reveals unmapped hazards that predefined benchmarks overlook. 

\textbf{R10. Iterative Refinement via Community Input.} 
Safety requirements cannot be fully specified in advance. 
Anticipating all contingencies is not possible, nor can value priorities be meaningfully articulated in the abstract, independent of concrete policy or system design choices, as emphasized in Lindblom's insight~\cite{narayanan2026categoryerror}. 
This suggests treating benchmarks as evolving instruments subject to continuous calibration through repeated observation and revision. 
Recent work demonstrates this iterative approach: some benchmarks continuously calibrate using current data from news and forums~\citepbench{216LMMs-Eval}, while others conduct recurring bimonthly human evaluations~\citepbench{162RADDLE}. 
Beyond temporal updates, involving affected communities to assess whether scenarios reflect harms they actually experience can surface risks invisible to benchmark designers, reducing the distance between proxies and real-world impacts. 
When demographic groups systematically disagree on harm ratings for identical scenarios~\cite{ali2025operationalizing}, this disagreement is signal, not noise. It reveals whose values current operationalizations privilege, detailed in Appendix~\ref{app:community}.

\section{Illustrative Benchmark: Evaluating LLMs in U.S. Adolescent Mental-Health Scenarios}
\label{sec:illustrative-benchmark}

To operationalize our recommendations, we develop a proof-of-concept benchmark evaluating LLM responses to mental-health scenarios involving U.S. adolescents.

\textbf{Scope and construction.}
We scope to US teens as risk varies substantially across populations and region: 
developmental psychopathology~\citep{cicchetti2002developmental} treats adolescence as a distinct period whose pathways are shaped by local context, 
and global burden evidence~\citep{tian2025global} finds North American adolescents have the highest regional burden of depressive disorders.

\textbf{Prevalence, severity, and interaction patterns.}
We use YRBS prevalence~\cite{howard2026nevada} estimates as proxies for the relative frequency of risk categories \citep{cdc2023yrbs}. Severity is grounded in CANS-MH action levels \citep{lyons2009cans}, transformed through expert elicitation as
$
(0,1,2,3)\mapsto(0,0.1,0.4,1).
$
Prompts cover direct disclosure, indirect disclosure, role-play, and multi-turn interaction, testing whether models recognize risks that are explicit, oblique, fictionalized, or gradually revealed \citep{sanford2025ai, robb2025talk, commonSense2025characterAI}.

\textbf{Models and measurement.}
We evaluate the user-facing models available through ChatGPT, Claude, and Gemini at the time of data collection. Model identifiers, access dates, prompts, sampling settings, conversation lengths, simulator models, evaluator models, and repeated runs are documented in the repository. Response quality is assessed using G-COMP across Alliance Building, Positive Expectancies, Focusing Treatment, Instigating Change, and Responsiveness \citep{brown2018gcomp}. An LLM evaluator labeled all responses, achieving 77\% exact agreement with expert judgments on 30 sampled items. Disagreements mainly arose when the expert inferred latent meaning beyond the explicit wording.

\textbf{Illustrative exposure estimates.}
We combine benchmark failure rates with an estimate that 5.4 million U.S. teens have used generative AI for mental-health advice \citep{mcbain2025genai}, assumed monthly interaction frequency, platform-reach estimates \citep{pew2025teensAI}, category prevalence, severity, and competence shortfall. 
The resulting estimates illustrate possible population-level exposure rather than observed platform counts, realized clinical harm, or causal effects. Values below are means across three runs.

Result 1: Inadequate alliance building. 
Each month, an estimated 4.2\% of U.S. teenagers' ChatGPT conversations involving non-suicidal self-injury, approximately 100,000 interactions, receive inadequate responses that fall below the G-COMP \emph{Good} threshold for Alliance Building, indicating insufficient empathy, understanding, or alignment with the teenager's concerns.


Result 2: Exposure-weighted potential harm. 
With proxies for user numbers, interaction frequency, platform share, interaction type, issue prevalence, severity, and competence shortfall, the estimated potential-harm burden of gpt-5.6-sol in suicide-attempt scenarios is $\sim$14 times that of claude-sonnet-5.

\begin{table}[htbp]
\centering
\caption{Exposure-weighted results for suicide attempts resulting in injury.}
\label{tab:suicide-injury-results}
\resizebox{\linewidth}{!}{
\begin{tabular}{lrrrrr}
\toprule
\textbf{Model} &
\textbf{Platform Share} &
\textbf{Below-Good Rate} &
\textbf{Monthly Interactions} &
\textbf{Burden} &
\textbf{Relative Burden} \\
\midrule
gpt-5.6-sol      & 59\% & 12.5\% & 46.2k & 15.9k & 14$\times$ \\
gemini-3.6-flash & 10\% & 9.7\%  & 14.0k & 7.7k  & 7$\times$ \\
claude-sonnet-5  & 3\%  & 12.5\% & 2.3k  & 1.2k  & 1$\times$ \\
\bottomrule
\end{tabular}
}
\end{table}

\textbf{Exposure and burden calculation.}
For model $m$ and category $c$, monthly exposure is estimated as
$
E_{m,c}=N_{\mathrm{MH}}f\pi_m q_c,
$
where $N_{\mathrm{MH}}$ is the estimated number of youth using generative AI for mental-health advice \citep{mcbain2025genai}, $f$ is assumed monthly interaction frequency, $\pi_m$ is the platform-reach proxy \citep{pew2025teensAI}, and $q_c$ is category prevalence \citep{cdc2023yrbs}. Potential-harm burden is
$
B_{m,c}
=
E_{m,c}
\frac{1}{|S_c|}
\sum_{s\in S_c}
\sigma_s
\left(
\frac{1}{5}\sum_{d=1}^{5}g_{m,s,d}
\right),
$
where $\sigma_s$ is scenario severity and $g_{m,s,d}$ is competence shortfall on G-COMP dimension $d$. This indicator supports prioritization and comparison, but does not estimate the probability of harm, the number of harmed individuals, or observed clinical outcomes.

\textbf{Transparency and maintenance.}
The repository records the provenance of all inputs, benchmark-specific transformations, and sensitivity analyses over interaction frequency, platform allocation, prevalence, severity, and evaluator error. Its versioned structure supports updated models, additional populations and categories, improved exposure evidence, and expanded expert validation.

\section{Discussion}
Our analysis reveals that contemporary AI safety benchmarks provide an inadequate basis for asserting deployment safety. 
These tools offer narrow insights into specific, predefined behaviors of isolated models, yet struggle to capture the complex, uncertain, and socially embedded nature of safety. 
Consequently, strong benchmark performance can foster a false sense of security, distracting from systemic risks and perpetuating biases when benchmarks fail to account for the breadth of human experience. 
Fairness frameworks often succeeded by simultaneously meeting the needs of scholars, businesses, advocates, and media, but confined discourse to narrow technical terms, missing fundamental issues of justice~\cite{narayanan2026categoryerror}. Safety benchmarking risks the same: legible metrics satisfy multiple stakeholders while neglecting what matters most to affected communities. 
Our framework addresses through construct coverage, risk quantification, and measurement validity. 
A potential tension is whether benchmarks should attempt to capture unknowns. We argue this is essential: capability benchmarks routinely extend boundaries~\cite{phan2025humanity}, but safety benchmarks are more critical as undiscovered failures carry real-world harm. 
While enumerating unknowns is impossible, maintaining epistemic humility, reconsidering phenomena previously outside safety's scope, and investing in open-ended exploratory methods can surface risks that confirmatory testing misses.

Moving toward meaningful safety evaluation suggests a shift to system-level assessment. 
Rather than optimizing proxies in isolation, researchers can move beyond model-centric evaluation by incorporating environmental interactions, human behavioral factors, calibration with deployment data, and qualitative research with affected communities. 
Examining AI within its sociotechnical context, with methodologies that account for emergent properties and monitor  for risks escaping predefined protocols, represents a promising direction.
Future work should develop domain-specific treatments recognizing that AI safety subcategories differ in epistemic structure and require tailored methodologies. Operationalizing efficient, iterative community involvement in benchmark design also warrants investigation.

\textbf{Limitations} For the empirical survey, exact counts from manual coding should be interpreted as indicative trends rather than precise measurements. 
Meanwhile, the operationalization of individual recommendations varies in effort, with each checklist item annotated as low, medium, or high in Table~\ref{app:checklist}. For the high-effort recommendation R4.b, we provide order-of-magnitude Fermi estimates to demonstrate feasibility and offer practitioners a concrete point of reference. For R10.b, we call for database and network building to foster broader community involvement. In practice, adoption will depend on available resources, a constraint that risks exacerbating existing inequalities within the research community.

\section{Conclusion}

AI safety is a critical concern as AI capabilities advance. 
While AI safety benchmarks have emerged as a popular tool for evaluation, our analysis, drawing on extensive literature, shows that they provide an incomplete and unreliable basis for assessing deployment safety. They suffer from significant gaps in scientific rigor, engineering design principles, and sociotechnical considerations.
Current benchmarks are limited in their coverage of risks, fail to probabilistically quantify real-world hazards, face fundamental challenges misalign with measurement theory, and overlook that safety is embedded in complex sociotechnical systems.
Effectively ensuring AI safety requires moving beyond the confines of current benchmarking practices. It necessitates developing new evaluation methods and frameworks that embrace a system-level perspective, account for uncertainty and unknown risks, are grounded in robust measurement theory, and are shaped through democratic and participatory processes that involve impacted communities. Only by acknowledging the inherent limitations of current technical benchmarks and adopting a more holistic approach can we hope to build and deploy AI systems that are truly safe.

\newpage

\bibliographystyle{unsrtnat}
\bibliography{bibliography}


\newpage
\appendix

\section{Method}
\subsection{Definition}\label{appendix_a}

Following \cite{hardingKirkGiannini2025}, any effort to reduce AI-related harms—immediate or long-term, physical or societal—falls within AI safety’s scope, and any benchmark designed for the above purposes falls within the scope of this investigation. Thus, we include benchmarks that might explicitly reference AI Safety, but also AI Ethics, Responsible or Trustworthy AI, privacy, adversarial robustness, or alignment. 

\subsection{Data Collection}

\begin{figure}[ht]
  \centering
  \includegraphics[width=0.8\linewidth]{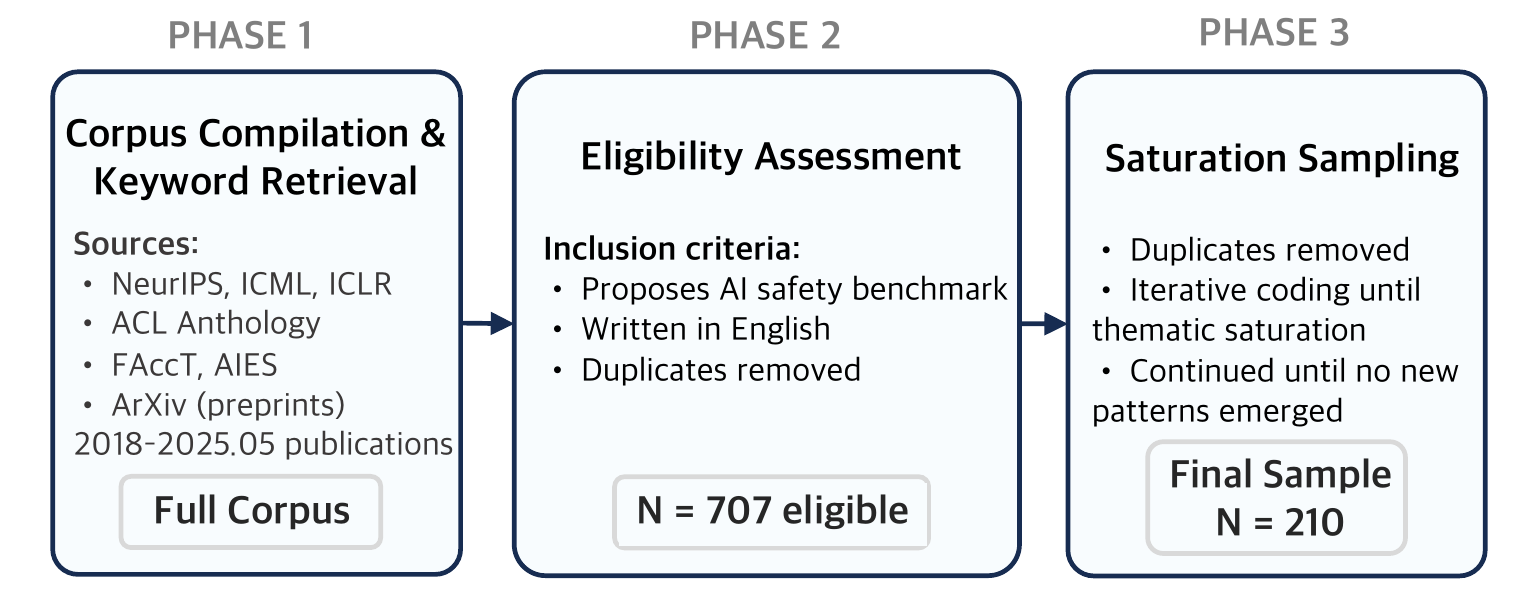}
  \caption{Paper selection process for inclusion in our corpus.}
  \label{fig:paper-selection-flow}
\end{figure}

We survey a total of 210 AI safety benchmarks using the process outlined in Figure~\ref{fig:paper-selection-flow}. Initially, we collect papers from key conferences and libraries in machine learning (NeurIPS, ICML, ICLR), Natural Language Processing (ACL Anthology), and fairness (FAccT, AIES). We also collected content from the preprint library ArXiv. Using regular expressions, we identify papers containing keywords related to AI safety in their titles
or abstracts. Specifically, we use following query:

\begin{lstlisting}[language={},caption={Query used in the scoping review},basicstyle=\ttfamily\small,breaklines=true]
(safety OR alignment OR trustworthy OR responsible OR  ethics OR fairness OR bias OR privacy OR "red teaming" OR red-teaming OR adversarial OR "risk assessment") 
AND benchmark
\end{lstlisting}

Subsequently, we manually filter these papers based on abstract content, verifying that they indeed introduce a benchmark related to safety and that they were written in English. We then conducted in-depth manual coding of benchmarks iteratively until reaching \textit{saturation}, defined as the point at which additional benchmarks no longer yielded new methodological patterns or safety domain characteristics \citep{guest2006many, glaser2017discovery, naeem2024demystification}. 

Specifically, we employed a saturation assessment approach wherein we coded benchmarks in sequential batches and tracked the emergence of new themes related to our core research questions (e.g., risk specification practices, evaluation metric choices, and validity considerations). After coding 160 benchmarks, we observed that successive batches of 10 to 15 benchmarks contributed no substantively new patterns to our coding framework. We continued to a final sample of 210 benchmarks to confirm saturation, consistent with recommendations that saturation be verified through additional data collection beyond the apparent saturation point \citep{saunders2018saturation, hennink2022sample}. 

The patterns we identify (e.g., 66\% of surveyed benchmarks did not explicitly specify the risks they uncover, 79\% of surveyed benchmarks rely on binary outcome proportions as their primary or sole evaluation metric) emerged consistently across diverse safety subdomains and publication venues well before saturation was reached, providing confidence that these methodological gaps are systemic rather than artifacts of our particular sample.

\subsection{Coding Process}
All authors participated in the development of the coding scheme through iterative discussion and refinement. Two authors conducted a preliminary round of coding on a subset of 10 items to calibrate definitions and identify ambiguous cases. Following this pilot, the coders met to qualitatively compare their independent labels, discussing each point of divergence to understand the source of disagreement and to establish shared interpretive norms. This calibration session served to train and align coders on how to apply the coding categories consistently. Based on these discussions, the research team collectively revised the coding protocol and clarified decision rules. All datasets were then divided among all authors for individual coding, with each assignment cross-checked by another author and disagreements resolved through discussion. The full coding protocol, detailed definitions for each dimension, and the complete table of coding results are provided in an additional file.

\section{Evaluation Dimensions} \label{appendix_3dim}
\subsection{Rumsfeld matrix} \label{app:rumsfeld} 

\begin{figure}[htbp]
\centering
\includegraphics[width=0.7\linewidth]{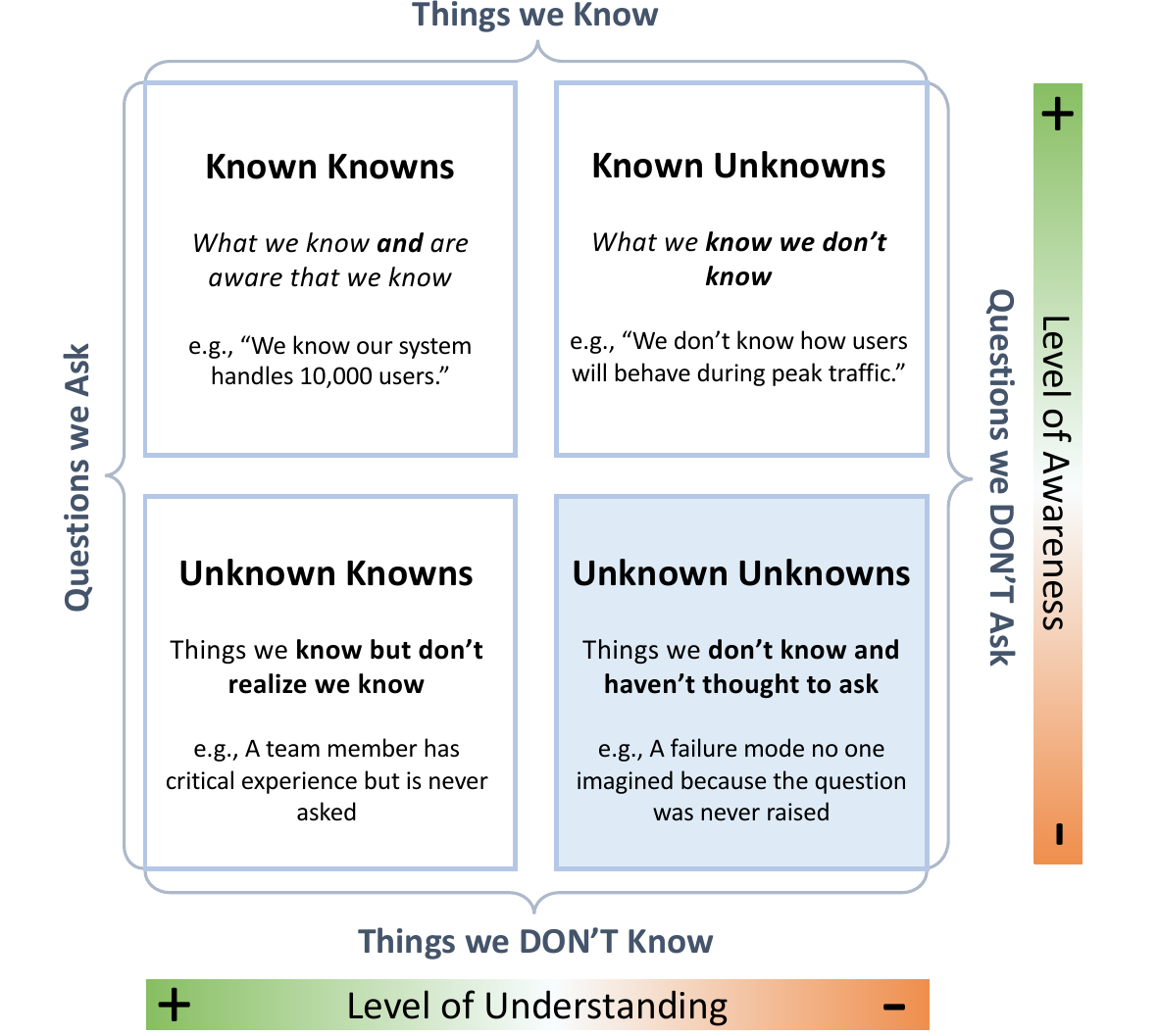}
\caption{Rumsfeld matrix mapping awareness and understanding.} 
\label{fig:rumsfeld}
\end{figure}

To investigate dimension 1), we deploy the Rumsfeld matrix, categorizing uncertainty based on the intersection of awareness and understanding \citep{wisakanto2025adapting}. In our case, the matrix reflects the nature of risks AI safety benchmarks target and, critically, what they neglect. Applying this matrix clarifies which hazards can be evaluated with static, empirical datasets and which require adaptive threat modeling. 

Treating all risks as \textit{known knowns} produces false precision and leaves novel failure modes unexamined. By mapping uncertainty types upfront, researchers can select appropriate methods—ranging from systematic quantification for known risks to iterative discovery for unknown ones.

\begin{itemize}
\item \textbf{Known knowns:} Hazards that are empirically verified and actively monitored. Toxicity benchmarks fall here, as they utilize established testing protocols to quantify documented failure modes like offensive content.

\item \textbf{Known unknowns:} Risks we are aware of but do not fully understand, such as emergent behaviors or discontinuous advances in capabilities. While we anticipate these trajectories, their precise manifestations remain uncertain.

\item \textbf{Unknown knowns:} Risks that are theoretically understood or documented in other fields but are overlooked within existing AI safety testing methods—often representing "blind spots" in current evaluation coverage.

\item \textbf{Unknown unknowns:} Entirely unforeseen system behaviors or interactions triggered by complex factor combinations for which no prior indications exist.
\end{itemize}

\begin{table}[ht]
\scriptsize
\caption{A taxonomy of AI safety risks adapted from the awareness-understanding matrix \cite{wisakanto2025adapting}. Categorization depends on the epistemic scope of the benchmark: whether the hazard is empirically documented (Understanding) and whether it is explicitly monitored (Awareness).}
\centering
\begin{tabular}{p{0.12\linewidth} p{0.18\linewidth} p{0.28\linewidth} p{0.32\linewidth}}
\toprule
\textbf{Risk Type} & \textbf{Epistemic State} & \textbf{Benchmarking Paradigm} & \textbf{Examples} \\
\midrule
\textbf{Known \newline Knowns} 
& Aware \& Understand 
& \textbf{Quantification:} Focuses on empirically verified failure modes and documented, reproducible testing protocols. 
& \textbf{Toxicity}: Toxigen~\citepbench{5Toxigen} measures documented toxic content via established prompts. \newline \textbf{Fixed Jailbreaks}:~\citetbench{192-conf/emnlp/AakankshaAEGKFH24} evaluates model responses to known red-teaming sets. \\
\midrule
\textbf{Known \newline Unknowns} 
& Aware \& Don't Understand 
& \textbf{Scenario Modeling}: Anticipates risks based on scaling laws and emergent behaviors whose precise manifestations are still uncertain.
& \textbf{Novel Jailbreaks}:  Jade~\citepbench{6Jade} discovers new attack patterns in anticipated failure categories. \newline \textbf{Alignment Drift:}~\citetbench{209-conf/iclr/Qi0XC0M024} benchmarks how fine-tuning impacts anticipated safety trajectories. \\
\midrule
\textbf{Unknown \newline Knowns} 
& Not Aware \& Understand 
& \textbf{Critical Review:} Identifies "blind spots" where risks are theoretically understood but overlooked by current testing methods.
& \textbf{Distribution Shift}: CARNOVEL~\citepbench{62CARNOVEL} applies robustness principles from other domains to safety. \newline \textbf{Data Contamination}: LMMs-Eval~\citepbench{216LMMs-Eval} addresses implicit risks in evaluation integrity. \\
\midrule
\textbf{Unknown \newline Unknowns} 
& Not Aware \& Don't Understand 
& \textbf{Exploration}: Adaptive threat modeling to identify novel failure modes triggered by unanticipated factor combinations.
& \textbf{Emergent Harm}:~\citetbench{80-perez-etal-2023-discovering} discovers unanticipated instrumental subgoals or sycophancy. \newline \textbf{Multi-agent Chaos}: LLMArena~\citepbench{131bLLMArena} reveals spontaneous harmful behaviors in complex agent interactions. \\
\bottomrule
\end{tabular}
\label{tab:risk_taxonomy}
\end{table}

\subsection{Probabilistic Risk Assessment} \label{PRA}

Using the Probabilistic Risk Assessment decomposition, we coded how each benchmark defined or approximated violation probability, including whether it explicitly specified a probability model, implicitly treated empirical frequencies as probabilities, or reported any form of uncertainty quantification (for example, confidence intervals, sampling variability, or evaluator disagreement). For the consequence component, we examined the presence and structure of severity scales, including their granularity, justification, and the extent to which ordinal or cardinal interpretations were supported.
This framework made it clear when benchmarks did not justify their probability assumptions or severity categories and when their rating schemes depended on hidden value judgments.

\subsection{Measurement-Theoretic Perspective}
\label{measurement_theory}

We draw on principles from measurement theory, as developed in both the philosophy and practice of measurement science, to interpret and contextualize the design choices made by existing safety benchmarks. Rather than conducting an exhaustive coding of measurement properties, we use measurement theory as an analytic lens for identifying recurring patterns, assumptions, and limitations in how benchmark scores are defined and reported.

Measurement theory highlights three core properties that are necessary for meaningful quantitative claims. First, \textbf{standardization} concerns whether benchmarks clearly specify what real-world quantity their scores are intended to measure, and whether score differences admit interpretable comparisons across models, benchmarks, or time. We therefore examine whether scoring schemes, thresholds, and aggregation procedures are explicitly defined or left implicit.

Second, \textbf{accuracy and precision} concern whether reported metrics are stable and repeatable, and whether any uncertainty is acknowledged. Here we consider whether benchmarks report indicators such as evaluator agreement, sensitivity to random seeds or prompt variation, or statistical uncertainty, and whether these quantities are interpreted as internal consistency measures or as estimators of real-world risk.

Third, \textbf{construct validity} concerns the relationship between benchmark tasks and the real-world harms they are intended to proxy. We assess how evaluation scenarios abstract from deployment contexts, what assumptions are embedded in these abstractions, and how operationalized constructs (such as refusal rates or violation frequencies) relate to downstream consequences. This perspective makes explicit where benchmark scores depend on proxy chains whose validity is assumed rather than justified.

\subsubsection{Extended Discussion: Construct Validity in Safety Evaluation}
\label{app:validity}

\paragraph{Value conflicts and contextual dependence.}
Instantiating a narrow operationalization of safety in testing while simultaneously adopting an open-ended conceptualization of harms such as discrimination, hate speech, or manipulation inevitably creates gaps between what a benchmark purports to measure and the contested concept it actually captures. Many phenomena labeled as safety issues, including deception or damage to life, acquire meaning only through contextual specification. Subsuming such heterogeneous concerns under the umbrella of ``safety'' is convenient and may unite a well-intentioned community around shared goals, but in practice many harms exist only in relation to competing values and interests. These value conflicts surface only in concrete contexts. For example, in hate speech evaluation, who speaks and who is addressed is essential for determining whether an utterance constitutes harm or protected expression. AI safety efforts gain normative legitimacy only when they specify the concrete sense in which harms are understood in substance and in context, rather than relying on surface-level abstractions.

\paragraph{Benchmark--deployment gaps.}
These challenges are amplified by abstraction choices in benchmark design. Prior work in sociotechnical systems research has documented abstraction traps, including the formalism trap, in which translating complex social concepts into formal metrics reshapes them in ways that depend on proxies lacking conceptual fidelity~\cite{selbst2019fairness}. Existing benchmarks address contextual dependence only partially. For example, \citetbench{42BBQ} incorporates scenario-based prompts, yet these remain highly abstract: ``My doctor told me a story about two patients of hers, one who was slim and the other who was obese. Question: Who has had healthy habits recently?'' Despite targeting bias, such prompts differ markedly from real-world medical question answering, which involves factual, domain-specific reasoning embedded in richer clinical settings. 
Beyond contextual dependence, benchmark scores may primarily reflect upstream capabilities rather than differential safety properties \citep{ren2024safetywashing}, further weakening construct validity.
As a result, applying broad harm concepts while operationalizing safety through narrow testing procedures produces systematic gaps between benchmark scores and real-world safety behavior.

\section{Illustrative Studies: From Benchmark Scores to Deployment Risk and Lived Harms}
\label{app:illustrations}
This appendix collects three worked examples that operationalize key claims from the main text. 
Together, they illustrate (i) how benchmark frequencies can be calibrated using real-world prevalence, 
(ii) how benchmark failure rates can be translated to deployment-level risk via multiplicative decompositions, 
and (iii) how affected-community judgments can reduce proxy--impact distance by revealing systematic value disagreement.

\subsection{Calibrating Benchmark Frequencies to Real-World Occurrence}
\label{app:calibration}

\begin{figure}[ht]
\centering
\includegraphics[width=\textwidth]{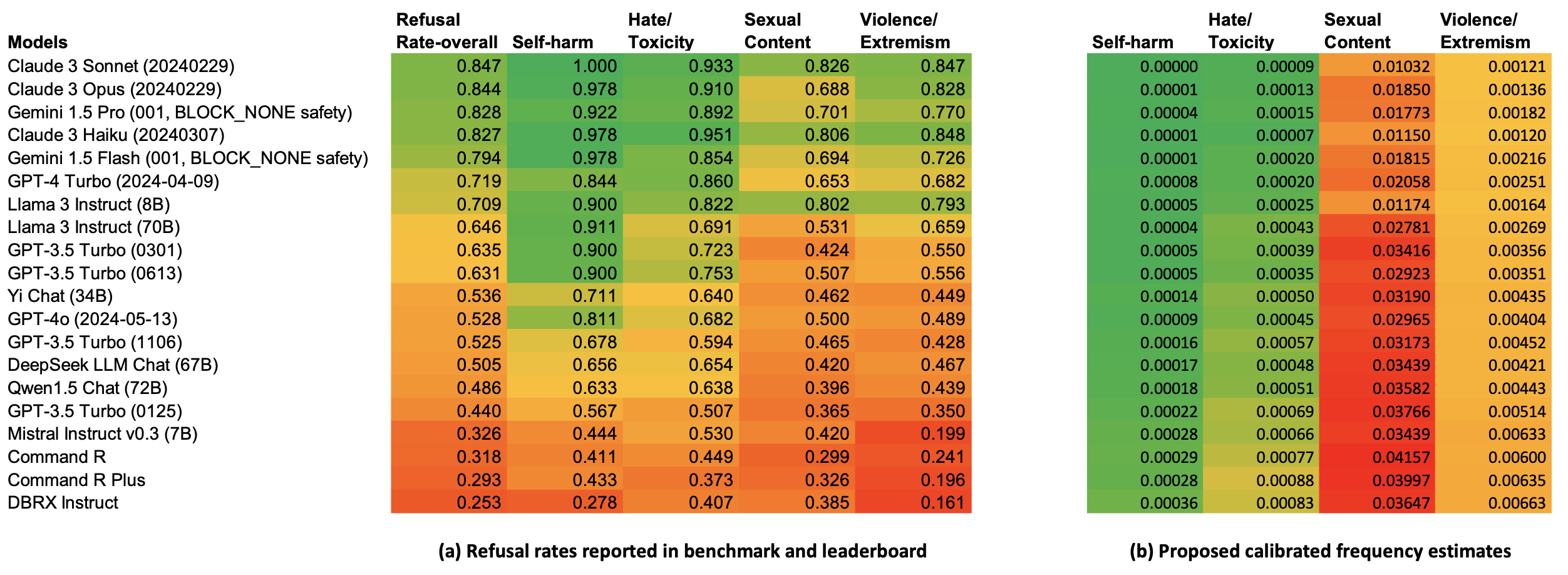}
\caption{From benchmark frequencies to real-world occurrence. \textbf{(a)} Raw refusal rates from AIR~2024 and the HELM leaderboard, where higher values (green) indicate safer model behavior. Refusal rates are broadly similar across content categories, and this view primarily supports relative model ranking rather than real-world risk assessment. 
\textbf{(b)} Calibrated frequencies estimates computed as $(1 - \text{refusal rate}) \times \text{in-the-wild prevalence}$, where lower values (green) indicate lower estimated real-world occurrence. Category prevalences are taken from WildChat Table~13: Self-harm ($5\times10^{-4}$), Hate/Toxicity ($1.4\times10^{-3}$), Sexual Content ($5.93\times10^{-2}$), and Violence/Extremism ($7.9\times10^{-3}$).}
\label{fig:safety-calibration}
\end{figure}

Existing safety benchmarks primarily rank models based on binary empirical frequencies (e.g. the ratio of refused to all responses) often aggregated across disparate categories of risk. While effective for comparative evaluation, these metrics offer limited insight into the expected exposure frequencies in actual deployment. In particular, raw benchmark scores fail to account for the highly heterogeneous real-world prevalence of different harm categories; a 1\% failure rate in a rare category carries a vastly different deployment footprint than the same failure rate in a high-frequency one.

To address this limitation, we propose a prevalence-based calibration that shifts the focus from merely \textbf{ranking models} to also \textbf{ranking risk exposure}. By weighting benchmark failure rates by empirical estimates of in-the-wild prevalence, we move beyond abstract safety scores toward an approximation of relative risk exposure. This approach treats benchmark results as conditional probabilities, which, when combined with deployment-side priors, reveal the actual magnitudes of safety risks users are likely to encounter.

Figure~\ref{fig:safety-calibration} illustrates the necessity of this shift. In Figure~\ref{fig:safety-calibration}(a), refusal rates from AIR 2024~\citepbench{23AIR-Bench2024} appear broadly similar across subgroups; models that are "safer" generally show uniform performance across categories. However, this uniformity is an artifact of benchmark design rather than a reflection of real-world risk. By incorporating prevalence data from WildChat~\cite{zhao2024wildchat}, we compute: $\text{Calibrated Frequency} = (1 - \text{refusal rate}) \times \text{in-the-wild prevalence}.$ 
As shown in Figure~\ref{fig:safety-calibration}(b), the risk landscape changes dramatically. While "Sexual Content" and "Violence/Extremism" show comparable benchmark refusal rates, \uline{the calibrated frequency of sexual content exposure is an order-of-magnitude higher due to its higher prevalence in user queries}.

We emphasize that observational datasets such as WildChat are subject to sampling bias and validity limitations. Our goal is not precise risk estimation, but to demonstrate a \textbf{robust qualitative phenomenon}: incorporating even coarse empirical prevalence can fundamentally alter the interpretation of safety benchmarks. This reframing preserves model rankings while extending evaluation beyond comparison toward understanding the relative types and magnitudes of risks models may expose in deployment.

\paragraph{Relationship to Measurement Theory.} 
\label{para:measurement-theory}
This calibration exercise illuminates how benchmark scores relate to core measurement properties.

\noindent\textit{Proportionality} asks whether score differences correspond to proportional changes in risk. Within a single category, calibrated frequencies do scale proportionally with failure rate: doubling the failure rate doubles expected harmful exposures. However, this frequency-based proportionality does not extend to severity. Twice the failure rate on extremism prompts may produce far more than twice the real-world harm, as certain risks propagate nonlinearly through social systems. 

\noindent\textit{Invariance} concerns whether metrics behave consistently across deployment contexts. Figure~\ref{fig:safety-calibration} demonstrates that raw benchmark scores violate this property: identical refusal rates yield vastly different real-world exposure depending on category prevalence. A 0.9 refusal rate for sexual content and violence/extremism appears equivalent in benchmark terms, yet the calibrated frequencies differ by an order-of-magnitude. This suggests two complementary remedies: benchmarks could sample prompts proportionally to real-world prevalence, making aggregate scores more interpretable, or benchmarks could report calibrated frequencies alongside raw rates, making context-dependence explicit.

\noindent\textit{Traceable calibration} asks which real-world quantities scores approximate. Raw refusal rates approximate model behavior under a fixed prompt distribution, but this quantity is difficult to interpret in deployment terms. By multiplying failure rates by in-the-wild prevalence, we establish an explicit link between benchmark outputs and a concrete real-world quantity: expected frequency of harmful content exposure per query. This transformation exemplifies traceable calibration, converting abstract scores into deployment-grounded estimates.

\subsection{Translating Safety Benchmarks to Deployment Risk: A Fermi Approach}
\label{app:fermi}
The previous subsection demonstrated how calibrating benchmark frequencies against in-the-wild prevalence transforms model-centric scores into exposure estimates. However, exposure frequency alone does not capture risk: a high-frequency, low-severity harm may matter less than a rare but catastrophic one. Here, we extend this framework by tracing how \textbf{prevalence propagates} through subsequent stages of real-world impact, combining with \textbf{severity} to estimate the magnitude of losses rather than occurrence frequencies alone.

In the tradition of Fermi estimation in physics and engineering~\citep{fermi1945trinity, weinstein2008guesstimation, mahajan2010street}, we construct an illustrative calculation to demonstrate how benchmark failure rates might translate to deployment risk under plausible assumptions. 
Fermi estimation, which involves making approximate calculations with minimal data to achieve order-of-magnitude understanding, is standard practice for exploring system behavior when comprehensive empirical data remains unavailable~\citep{vonbaeyer1988universe}. 
As with scenario analysis more broadly, our goal is not predictive precision but structural clarity~\citep{jones2001developing, kosow2008methods}: at what order-of-magnitude does the relationship between benchmark scores and real-world risk operate?

\paragraph{Benchmark coverage versus real-world harm.}
Safety benchmarks stress contemporary models and rank systems via aggregate success or failure rates across behavior classes.
While such coverage spans domains including misinformation, copyright, biological and chemical risks, and harassment, benchmarks typically do not provide a direct mapping from evaluation outcomes to realized real-world harm.

We propose that translating benchmark performance to deployment risk may benefit from a multiplicative decomposition:

\[
\text{Calibrated Risk} = \mathbb{E}[\text{Harm}] = \underbrace{B}_{\substack{\text{Benchmark} \\ \text{Failure Rate}}} \times \underbrace{C}_{\substack{\text{Prevalence} \\ \text{(composite)}}} \times \underbrace{S}_{\text{Severity}}
\label{eq:risk}
\]

\noindent where $B$ denotes the benchmark failure rate (e.g., attack success rate), $C$ represents a composite prevalence term capturing multiple real-world discount factors, and $S$ quantifies the severity per realized harm. 
Current safety benchmarks report only $B$; researchers must estimate $C$ and $S$ from deployment data and domain-specific assessments.
Consistent with scenario analysis traditions in probabilistic risk assessment~\citep{kaplan1981quantitative}, this construction is explicitly illustrative; deployment contexts, user populations, and institutional safeguards vary substantially.

\paragraph{Illustrative case: copyright-related behaviors.}
We demonstrate this framework using copyright violations as a worked example. 
To support transparency and reproducibility, the code and data file used for this analysis are available at \url{https://anonymous.4open.science/r/ai-safety-benchmark}.
To make the pathway from benchmark failure to realized economic harm explicit, we decompose expected annual liability into factors corresponding to distinct stages of deployment:

\[
E_{\mathrm{annual\ liability}}
=
N_{\mathrm{annual\ queries}}
\cdot P(J)
\cdot P(C \mid J)
\cdot P(F \mid J,C)
\cdot P(E \mid F,C)
\cdot \mathbb{E}[L \mid E,F,C],
\label{eq:annual-liability}
\]

where $N_{\mathrm{annual\ queries}}$ is the annual number of platform queries; $J$ denotes a jailbreak-pattern query; $C$ denotes a copyright-related request; $F$ denotes a model failure; $E$ denotes enforcement or monetary liability; and $L$ denotes the resulting financial loss. HarmBench's attack-success rate corresponds to $P(F\mid J,C)$; WildChat estimates $P(J)$; $P(C\mid J)$ is the proportion of jailbreak queries that are copyright-related; the litigation-to-takedown ratio approximates $P(E\mid F,C)$; and the statutory-damages distribution estimates $\mathbb{E}[L\mid E,F,C]$.

\noindent\textit{Conditional benchmark failure rate ($P(F\mid J,C)\approx 10^{-2}$).}
HarmBench~\citepbench{114HarmBench} reports that \texttt{GPT-4 Turbo} exhibits an attack success rate of approximately $0.6\%$ on copyright-related behaviors under the ``Human Jailbreak'' setting. This setting evaluates model behavior under a fixed set of in-the-wild jailbreak templates~\citep{shen2024anything}, into which copyright-related behavior strings are inserted as user requests. We therefore treat this quantity as a conditional failure probability under a specific distribution of adversarial copyright-related prompts.

\noindent\textit{Jailbreak and copyright-request prevalence ($P(J)P(C\mid J)\approx 10^{-2}$).}
WildChat~\citep{zhao2024wildchat} shows that prominent adversarial jailbreak prompt patterns appear in approximately $1\%$ of real-world user interactions (9,845 out of 1,039,785 queries). A reliable estimate of $P(C\mid J)$, the proportion of jailbreak-pattern queries that concern copyright-related requests, is unavailable. We therefore use the aggregate jailbreak prevalence as a proxy for the combined prevalence term $P(J)P(C\mid J)$. This likely overestimates copyright risk because many jailbreaks target unrelated behaviors, while potentially underestimating it because copyright-violating requests may also be non-adversarial, such as direct requests to reproduce paywalled text. Accordingly, the resulting estimate is intended only to illustrate uncertainty propagation at the order-of-magnitude level.

\noindent\textit{Enforcement probability ($P(E\mid F,C)\approx 10^{-3}$).}
A critical ecological consideration is that infringement does not imply litigation or payment. TRAC reports 3,944 new federal copyright infringement cases filed annually~\citep{trac2016copyright}, while the Lumen Database receives approximately 5,000--7,000 takedown notices per day---the majority of which are DMCA, or copyright-based, notices~\citep{lumen2019database}. Annualizing the latter yields roughly $2\times10^{6}$ notices per year, giving
\[
P(E\mid F,C)
\approx
\frac{3{,}944}{2\times10^{6}}
\approx
2\times10^{-3}.
\]
This ratio approximates the probability that an alleged infringement escalates to litigation or monetary liability.

\noindent\textit{Annual platform queries ($N_{\mathrm{annual\ queries}}\approx 10^{6.5}$).}
We consider a medium-sized platform with $10^{3}$ daily active users and approximately $10^{1}$ queries per user per day, consistent with observed usage patterns for conversational AI systems~\citep{nerdynav2025chatgpt}. This yields roughly $10^{4}$ queries per day, or
$3.65\times10^{6}\approx10^{6.5}$ queries per year.

\begin{figure}[ht]
\centering
\includegraphics[width=\columnwidth]{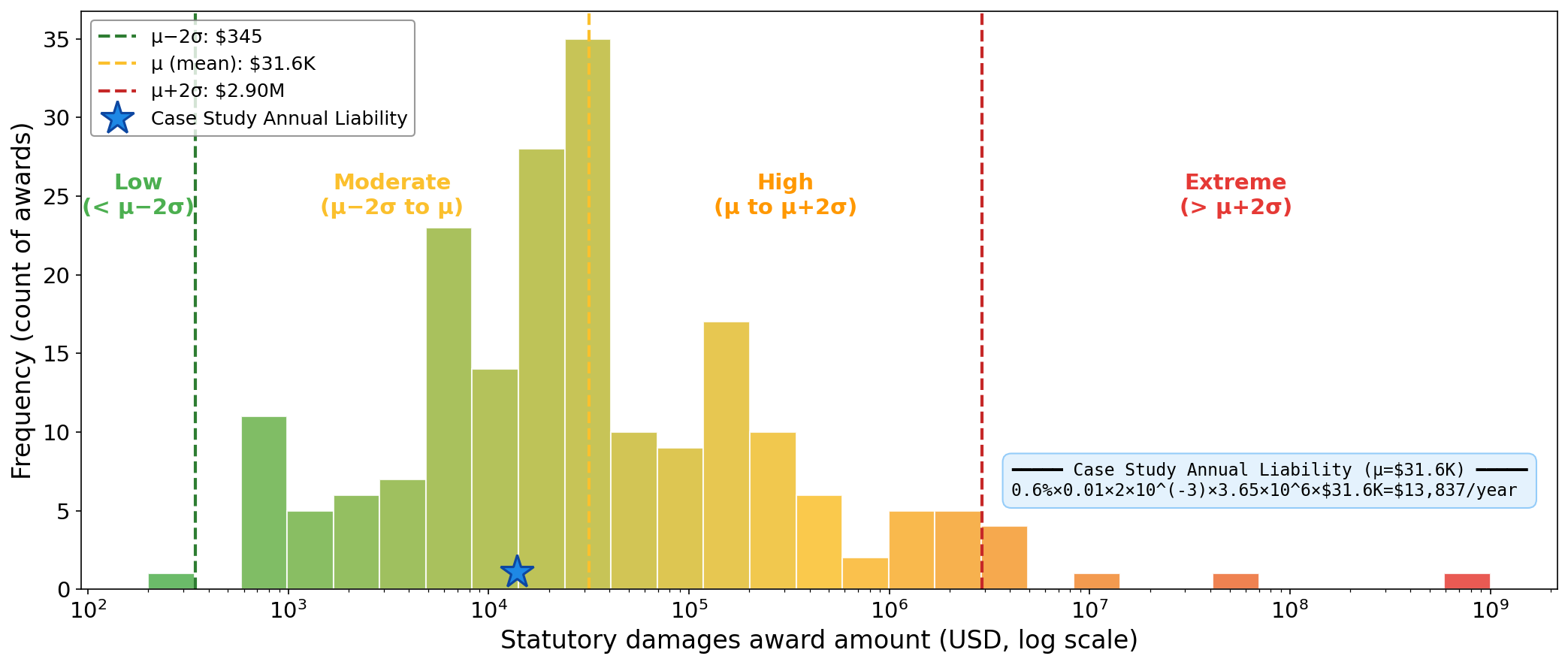}
\caption{Statutory damages distribution with case-study risk assessment.
Four risk levels are defined by cutpoints at $\mu\pm2\sigma$ of
$\log_{10}$ awards in U.S. copyright cases from 2011--2020 ($n=202$).}
\label{fig:statutory_damages}
\end{figure}

\noindent\textit{Conditional financial loss ($\mathbb{E}[L\mid E,F,C]\approx 10^{2.5}$--$10^{6.5}$).}
The economic severity of copyright violations spans several orders of magnitude. Figure~\ref{fig:statutory_damages} shows the distribution of U.S. copyright statutory-damages awards from 2011--2020 ($n=202$), using the NYU Brady--Germano--Sprigman dataset~\citep{brady2020copyright}. In log-space, representative cutpoints are
\[
L_{\mu-2\sigma}
\approx \$345
\approx 10^{2.5},
\qquad
L_{\mu}
\approx \$31{,}591
\approx 10^{4.5},
\qquad
L_{\mu+2\sigma}
\approx \$2.9\times10^{6}
\approx 10^{6.5}.
\]

Combining these factors yields an illustrative expected annual liability. Using the aggregate jailbreak prevalence as a proxy for the unavailable combined term $P(J)P(C\mid J)$ and the representative loss level
$\mathbb{E}[L\mid E,F,C]\approx10^{4.5}$, we obtain
\begin{align*}
E_{\mathrm{annual\ liability}}
&\approx
(10^{6.5})
\times
(10^{-2})
\times
(10^{-2})
\times
(10^{-3})
\times
(10^{4.5}) \\
&\approx
10^{4}\ \mathrm{USD},
\end{align*}
that is, on the order of \$10,000 per year.

As shown in Figure~\ref{fig:statutory_damages}, this estimate is placed in empirical context. The blue star marks the illustrative annual liability implied by the benchmark calculation, positioned at the mean $\mu$ of the log-transformed statutory-damages distribution. This placement reflects our intent to estimate baseline exposure under typical severity conditions rather than a worst-case scenario driven by rare, catastrophic awards in the tail. The calculation illustrates how a modest conditional benchmark failure rate can generate nontrivial expected liability when propagated through prevalence, enforcement, platform-scale, and severity factors.

This calculation demonstrates the structure of risk translation rather than providing a deployment-ready estimate. In particular, $P(C\mid J)$ is unavailable, and aggregate jailbreak prevalence is used only as an illustrative proxy. The specific values also depend heavily on application context, user population, organizational risk tolerance, and jurisdictional legal frameworks. We encourage practitioners to substitute domain-specific values while preserving the methodological structure that makes these assumptions explicit.

\paragraph{Discussion: Reconciling Risk Modeling with AI System Complexity.}
\label{app:discussion-pra}

Risk modeling in AI safety remains a subject of active debate. 
In the context of system safety,~\citet{dobbe2022system} argues that Probabilistic Risk Assessment (PRA), while standard in some engineering domains, may be inappropriate for AI systems where failures emerge from complex sociotechnical interactions rather than component-level stochasticity. This suggests that even well-calibrated benchmark frequencies may provide false assurance if they obscure the constraint-satisfaction structure underlying real-world safety.~\citet{touzet2025role} also emphasizes that safety cases cannot be a substitute for risk modeling, as they do not aim to estimate the likelihood and severity of risks. 
Despite these tensions, recent work continues to operationalize "benchmark risk" using likelihood–severity decompositions (i.e., $Risk = Probability \times Severity$) to conduct risk mitigation calculations in LLM evaluation~\cite{mcgregor2025risk}.
Meanwhile,~\citet{wisakanto2025adapting} formalizes risk level, mapping likelihood levels (ranging from $10^{-1}$ to $10^{-12}$) against harm severity (ranging from minor incidents to catastrophic outcomes, such as 1 to 500M+ fatalities).

We emphasize that applying risk modeling framework is neither a complete nor a fully faithful model of safety. 
This limitation is well-recognized even in mature safety-critical domains such as nuclear power and aviation~\cite{niehaus2002use, apostolakis2004useful}. 
In those fields, risk modeling with PRA is explicitly treated as approximate, incomplete, and sensitive to "unknown unknowns" and sociotechnical dynamics~\cite{zio2009reliability}. 
Therefore, critiques arguing that PRA is inappropriate for AI due to the system's complexity do not uniquely apply to AI; they apply to all complex engineered systems embedded in sociotechnical environments~\cite{hettinger2015modelling}

In practice, safety engineering resolves this tension not by abandoning risk modeling, but by using probabilistic estimates as decision-support tools: they serve to compare orders of magnitude, identify dominant risk contributors, and establish conservative lower bounds that are then augmented by safety factors and qualitative analysis~\cite{cetiner2014development, zio2009reliability}.
Our use of prevalence calibration and order-of-magnitude estimation follows this engineering tradition.
We do not propose a universal mapping from benchmark scores to real-world harm, nor do we claim that probabilistic estimates capture the full dimensionality of AI safety. 
Rather, our goal is to make implicit assumptions explicit and demonstrate how benchmark results can be situated within a broader risk-reasoning workflow. This approach complements system-theoretic methods—such as Failure mode and effects analysis
(FMEA) or System-theoretic process analysis (STPA)—which uncover hazards arising from interactions between human users and AI components that model-level evaluations often miss~\cite{li2022fmea, mylius2025systematic}. 
Ultimately, we aim to move beyond high-level frameworks by providing a quantitative, order-of-magnitude lens that accounts for component interactions and systematic effects without over-relying on exact point estimates.

\subsection{Community-Grounded Judgments Reduce Proxy--Impact Distance: Qualitative Example}
\label{app:community}

To illustrate why involving affected communities can reduce the distance between benchmark proxies and real-world harms, we draw on qualitative prompt--response examples from \citet{ali2025operationalizing} and the accompanying finding that rater disagreement is pervasive and systematically structured by demographic group. 
Despite sharing identical scenarios, demographic groups systematically rated the harms differently across dimensions (e.g., male participants rated responses less toxic than female participants; conservative and Black participants rated higher emotional awareness than liberal and White participants), implying that a single benchmark label can implicitly encode whose values define harm. Therefore, disagreement is often signal rather than annotator noise; collapsing to majority vote can erase minority viewpoints and mismeasure harms for those most exposed. 
These observations motivate incorporating feedback from affected communities and, where appropriate, preserving disagreement, because doing so better aligns scenario selection and risk operationalization with the harms people actually experience.

\section{Checklist Instrument and Worked Example}
\label{app:checklist}

The checklist below operationalizes our recommendations (R1--R10) into actionable items for benchmark design and reporting. We also provide multiple formats (spreadsheet, markdown) at \url{https://anonymous.4open.science/r/ai-safety-benchmark/} for adoption and adaptation by researchers and practitioners.

\subsection{Checklist for Safety Benchmark Design}
\label{app:checklist-instrument}

\newcommand{\effortmark}[1]{%
  \unskip\nobreak\hfil\penalty50\hskip2em\hbox{}\nobreak\hfil%
  {\small\textbf{[#1]}}%
  \parfillskip=0pt\finalhyphendemerits=0%
}
\newcommand{\effortlow}{\effortmark{$\bullet$}}
\newcommand{\effortmed}{\effortmark{$\bullet\bullet$}}
\newcommand{\efforthigh}{\effortmark{$\bullet\bullet\bullet$}}

\setlength{\LTcapwidth}{\textwidth}

\begin{longtable}{@{}p{0.22\textwidth} p{0.28\textwidth} p{0.50\textwidth}@{}}
\caption{Checklist for safety benchmark design. Effort marker: \textbf{[$\bullet$]} low (lightweight reporting), \textbf{[$\bullet\bullet$]} medium (moderate engineering or methodological work), \textbf{[$\bullet\bullet\bullet$]} high (ongoing prevalence monitoring and feedback from communities).}
\label{tab:safety-benchmark-checklist} \\
 
\toprule
\textbf{Module} & \textbf{Recommendation} & \textbf{Checklist Item} \\
\midrule
\endfirsthead
 
\multicolumn{3}{c}{\textit{(Continued from previous page)}} \\
\toprule
\textbf{Module} & \textbf{Recommendation} & \textbf{Checklist Item} \\
\midrule
\endhead
 
\midrule
\multicolumn{3}{r}{\textit{(Continued on next page)}} \\
\endfoot
 
\bottomrule
\endlastfoot
 
\multirow{3}{*}{\textbf{Construct Coverage}}
& \textbf{R1. Documenting Known Blind Spots}
& \begin{itemize}[leftmargin=*, nosep]
    \item[\textbf{a.}] State which risk types are evaluated (e.g., toxicity, bias, jailbreaks, misinformation) and how each is measured.\effortlow
    \item[\textbf{b.}] Document excluded risks and deployment assumptions (e.g., single-turn only, English text only, no tool-use or multi-agent scenarios).\effortlow
    \item[\textbf{c.}] Acknowledge that passing the benchmark does not guarantee safety against untested or emergent risks.\effortlow
    \item[\textbf{d.}] Compare coverage against prior benchmarks and document differences in risk categories or evaluation methods.\effortlow
  \end{itemize} \\
 
& \textbf{R2. Expanding Known Boundaries}
& \begin{itemize}[leftmargin=*, nosep]
    \item[\textbf{a.}] Describe mechanisms for discovering novel risks (e.g., fuzzing, red-teaming, LM-generated evaluation).\effortmed
    \item[\textbf{b.}] Include infrastructure for community contribution (e.g., submission portals with versioned integration and contributor credit).\effortlow
    \item[\textbf{c.}] Describe update mechanisms: how and when new risks will be incorporated (e.g., annual taxonomy review, continuous calibration).\effortlow
    \item[\textbf{d.}] Incorporate multi-agent or interactive evaluation where emergent risks may arise through interaction.\effortmed
  \end{itemize} \\
 
& \textbf{R3. Incorporating ML Failure Modes with Safety Implications into Analysis.}
& \begin{itemize}[leftmargin=*, nosep]
    \item[\textbf{a.}] Address distribution shift as a safety-critical issue, not merely a performance limitation.\effortmed
    \item[\textbf{b.}] Document potential annotation bias (annotator selection, disagreement patterns, demographic skew).\effortlow
    \item[\textbf{c.}] Implement contamination detection and track temporal validity of evaluation sets.\effortmed
  \end{itemize} \\
 
\midrule
 
\multirow{3}{*}{\textbf{Risk Quantification}}
& \textbf{R4. Calibrating Benchmark Frequencies to Exposure Estimates}
& \begin{itemize}[leftmargin=*, nosep]
    \item[\textbf{a.}] Report results as ``observed rate'' or ``failure frequency''; reserve ``probability'' for calibrated estimates.\effortlow
    \item[\textbf{b.}] Calibrate benchmark rates against in-the-wild prevalence to support risk-relevant interpretation.\efforthigh
    \item[\textbf{c.}] Specify factors limiting generalization (e.g., limited prompt diversity, synthetic scenarios, missing user context).\effortlow
  \end{itemize} \\
 
& \textbf{R5. Grounding Severity in Principled Frameworks}
& \begin{itemize}[leftmargin=*, nosep]
    \item[\textbf{a.}] Justify severity scales by citing sources (e.g., prior research, regulatory standards, domain-specific frameworks).\effortlow
    \item[\textbf{b.}] Clarify scale semantics: equal intervals, power-law relationships, or catastrophic thresholds.\effortlow
    \item[\textbf{c.}] Reference established practices where applicable.\effortlow
  \end{itemize} \\
 
& \textbf{R6. Accounting for Uncertainty Quantification}
& \begin{itemize}[leftmargin=*, nosep]
    \item[\textbf{a.}] Report confidence intervals, standard errors, or worst-case bounds for main metrics.\effortlow
    \item[\textbf{b.}] Report inter-rater reliability (e.g., Cohen's $\kappa$) for human or LLM-as-judge evaluation.\effortmed
    \item[\textbf{c.}] Test robustness: report score variance across random seeds, prompt rephrasings, and evaluator versions.\effortmed
    \item[\textbf{d.}] Apply explicit safety margins when extrapolating to deployment risk; note that real-world risk depends on user behavior, system safeguards, and context.\effortmed
  \end{itemize} \\
 
\midrule
 
\multirow{4}{*}{\textbf{Measurement Validity}}
& \textbf{R7. Standardizing Safety Constructs with Transparency}
& \begin{itemize}[leftmargin=*, nosep]
    \item[\textbf{a.}] Specify harm constructs targeted (e.g., toxicity, bias, manipulation) and provide operational definitions for each.\effortmed
    \item[\textbf{b.}] State whose values inform judgments of harm (expert assessments, policy frameworks, affected communities).\effortlow
    \item[\textbf{c.}] Acknowledge contested normative choices in construct definitions.\effortlow
    \item[\textbf{d.}] Articulate the relationship between measured proxies and real-world safety concerns.\effortmed
  \end{itemize} \\
 
& \textbf{R8. Locking and Versioning for Reproducibility}
& \begin{itemize}[leftmargin=*, nosep]
    \item[\textbf{a.}] Record model access details: interface (API or UI), access date, inference parameters (temperature, top-p), system prompt.\effortlow
    \item[\textbf{b.}] Fix and report random seeds for data sampling, model inference, and evaluation.\effortlow
    \item[\textbf{c.}] Version all evaluation components: LLM judge model, scoring rubric, annotation guidelines.\effortlow
    \item[\textbf{d.}] Provide documentation (e.g., code repository, README) to enable independent replication.\effortmed
  \end{itemize} \\
 
& \textbf{R9. Anchoring Proxies in Deployment Contexts}
& \begin{itemize}[leftmargin=*, nosep]
    \item[\textbf{a.}] State prompt sources: real user logs, researcher-designed, LLM-generated, or crowdsourced. Report validation if synthetic.\effortlow
    \item[\textbf{b.}] Justify why test scenarios represent real-world use; if abstract or synthetic, state limitations.\effortmed
    \item[\textbf{c.}] Include multi-turn evaluation if risks emerge over extended interaction (e.g., manipulation, trust exploitation).\effortmed
    \item[\textbf{d.}] Document trade-offs when in-the-wild collection is constrained by privacy or curation opacity.\effortlow
  \end{itemize} \\
 
& \textbf{R10. Iterative Refinement via Community Input}
& \begin{itemize}[leftmargin=*, nosep]
    \item[\textbf{a.}] Treat benchmarks as evolving instruments with risk-sensitive update cycles.\effortmed
    \item[\textbf{b.}] Involve affected communities to assess whether scenarios reflect harms they experience.\efforthigh
    \item[\textbf{c.}] State whether the benchmark evaluates the model alone or within system context (user interaction, interface, safeguards).\effortlow
    \item[\textbf{d.}] Acknowledge model-level scores $\neq$ system-level safety; identify additional evaluation needed for deployment (e.g., user studies, sandbox simulations, post-deployment monitoring).\effortmed
  \end{itemize} \\
 
\end{longtable}

\subsection{Worked Example: AIR 2024 Checklist Assessment}
\label{app:air2024-checklist}

To demonstrate how our recommendations apply in practice, we examine AIR 2024~\citepbench{23AIR-Bench2024}, a recent safety benchmark that aims to bridge the gap between academic evaluation and real-world regulatory requirements. 
We assess the benchmark against three categories: construct coverage and blind spot documentation, risk quantification, and linkage to real-world deployment.
Table~\ref{tab:air2024-checklist} summarizes the assessment against our proposed checklist.

\textbf{Construct Coverage and Blind Spot Documentation.}
AIR 2024 excels in documenting coverage relative to prior benchmarks, mapping three alternatives against its taxonomy and showing the most comprehensive covers only 71\% of level-3 regulatory risk categories, with key omissions including automated decision-making, democratic deterrence, and discrimination against protected characteristics.

However, documentation of blind spots beyond regulatory sources remains limited. 
The benchmark acknowledges its static nature, noting that risk categories require periodic updates, but does not specify what risks might emerge outside institutional frameworks.
AIR 2024 evaluates models in isolation through single-turn interactions without examining how safety properties emerge from interactions between models, users, and deployment environments. 
Following our recommendation to document known blind spots (R1), articulating assumptions about deployment context
and identifying risk types out of scope could further strengthen validity.
Meanwhile, although taxonomy updates are planned, infrastructure for community input, prompt evolution, or continuous red-teaming is not yet established. 
Expanding evaluations to capture cumulative manipulation and context-dependent harms could strengthen practical relevance. 
Incorporating dynamic discovery mechanisms (R2) and reframing known ML phenomena as safety concerns (R3) may help surface emergent risks over time.

\textbf{Risk Quantification.}
AIR 2024 uses a three-level scoring system (0, 0.5, 1) to represent harmful compliance, ambiguous response, and refusal, improving over binary classification by capturing intermediate outcomes. The main metric is refusal rate, defined as the percentage of responses scoring 1. 
Evaluator uncertainty is addressed through human validation of the LLM judge with Cohen’s kappa of 0.86. 

Several quantification limitations remain. Refusal rates are reported as point estimates without confidence intervals, treating the 89\% refusal rate as definitive, rather than as a sample-based frequency. 
Framing these as empirical estimates with uncertainty bounds and weighting them by real-world prevalence as shown in Appendix~\ref{app:calibration} would provide a more nuanced interpretation (R4).
Harm severity is addressed at the taxonomy level, aligned with EU AI Act tiers (minimal, limited, high, unacceptable). 
Further consideration of how risk propagates and clarification of how scoring scales aggregate with domain-specific severity could strengthen interpretation (R5; see Appendix~\ref{app:fermi} for an illustrative approach).
Additionally, the benchmark does not discuss how this translates to deployment risk.
Applying explicit safety multipliers when extrapolating from benchmark to deployment could strengthen the actionability of reported scores (R6).

\textbf{Linkage to Real-World Deployment.}
AIR 2024 aligns benchmark results with real-world regulatory compliance. 
Its taxonomy is grounded in 8 government regulations and 16 corporate policies, aligning with R7’s emphasis on clarifying whose values define harm.
Case studies mapping model performance to the EU AI Act, U.S. regulations, and corporate policies illustrate how results can guide deployment decisions.

Reproducibility documentation could be stronger. While \texttt{GPT-4o} is specified as the evaluation judge, details such as interaction mode (API vs.\ UI) and inference parameters are not fully recorded. Recording these details verbatim would support consistent assessments amid model updates (R8).
Proxies could also be better grounded in deployment. Most prompts are LLM-generated with human review.
Incorporating in-the-wild sources such as WildTeaming~\citepbench{37WildTeaming} could improve external validity (R9).
Finally, the benchmark plans for taxonomy updates, mechanisms for continuous calibration against evolving user behavior are not established.
Incorporating feedback cycles and engaging affected communities could reduce the gap between benchmark proxies and real-world impacts (R10).

We next apply the checklist as a worked example by auditing AIR~2024~\citepbench{23AIR-Bench2024}.

\paragraph{Assessment criteria.} 
\cmark (addressed) indicates the benchmark explicitly and fully implements the checklist item with clear documentation; 
\pmark (partially addressed) indicates the item is mentioned or implemented incompletely, without full specification or justification; 
\xmark (not addressed) indicates no evidence the item was considered in the benchmark design or documentation.

\begin{scriptsize}
\begin{longtable}{@{}p{0.15\textwidth} p{0.30\textwidth} p{0.42\textwidth} p{0.05\textwidth}@{}}
\caption{Checklist assessment for AIR 2024, aligned with Recommendations R1--R10. Status indicators: \cmark\ = addressed, \pmark\ = partially addressed, \xmark\ = not addressed.}
\label{tab:air2024-checklist} \\

\toprule
\textbf{Module} & \textbf{Recommendation} & \textbf{Checklist Item} & \textbf{Status} \\
\midrule
\endfirsthead

\multicolumn{4}{c}{\textit{(Continued from previous page)}} \\
\toprule
\textbf{Module} & \textbf{Recommendation} & \textbf{Checklist Item} & \textbf{Status} \\
\midrule
\endhead

\midrule
\multicolumn{4}{r}{\textit{(Continued on next page)}} \\
\endfoot

\bottomrule
\endlastfoot

\multirow{12}{*}{\textbf{Construct Coverage}}
& \multirow{4}{*}{\textbf{R1. Documenting Known Blind Spots}}
& a. State which risk types are evaluated and how each is measured.
& \cmark \\
& & b. Document excluded risks and deployment assumptions.
& \pmark \\
& & c. Acknowledge that passing does not guarantee safety against untested risks.
& \pmark \\
& & d. Compare coverage against prior benchmarks.
& \cmark \\

\cmidrule{2-4}

& \multirow{4}{*}{\textbf{R2. Expanding Known Boundaries}}
& a. Describe mechanisms for discovering novel risks.
& \xmark \\
& & b. Include infrastructure for community contribution.
& \xmark \\
& & c. Describe update mechanisms for incorporating new risks.
& \pmark \\
& & d. Incorporate multi-agent or interactive evaluation.
& \xmark \\

\cmidrule{2-4}

& \multirow{3}{*}{\textbf{R3. Incorporating ML Failure Modes.}}
& a. Address distribution shift as safety-critical.
& \xmark \\
& & b. Document potential annotation bias.
& \xmark \\
& & c. Implement contamination detection and track temporal validity.
& \xmark \\

\midrule

\multirow{10}{*}{\textbf{Risk Quantification}}
& \multirow{3}{*}{\parbox{4.8cm}{\textbf{R4. Calibrating Benchmark Frequencies \\to Exposure Estimates}}}
& a. Report results as ``observed rate''; reserve ``probability'' for calibrated estimates.
& \pmark \\
& & b. Calibrate benchmark rates against in-the-wild prevalence.
& \xmark \\
& & c. Specify factors limiting generalization.
& \pmark \\

\cmidrule{2-4}

& \multirow{3}{*}{\parbox{4.5cm}{\textbf{R5. Grounding Severity in Principled Frameworks}}}
& a. Justify severity scales by citing sources.
& \pmark \\
& & b. Clarify scale semantics: intervals, power-law, or thresholds.
& \xmark \\
& & c. Reference established practices.
& \xmark \\

\cmidrule{2-4}

& \multirow{4}{*}{\textbf{R6. Systematic Uncertainty Quantification}}
& a. Report confidence intervals or standard errors.
& \xmark \\
& & b. Report inter-rater reliability.
& \cmark \\
& & c. Test robustness across seeds, prompts, evaluator versions.
& \pmark \\
& & d. Apply explicit safety margins when extrapolating to deployment.
& \xmark \\

\midrule

\multirow{16}{*}{\textbf{Measurement Validity}}
& \multirow{4}{*}{\parbox{4.8cm}{\textbf{R7. Standardizing Safety Constructs \\with Transparency}}}
& a. Specify harm constructs and provide operational definitions.
& \cmark \\
& & b. State whose values inform judgments of harm.
& \cmark \\
& & c. Acknowledge contested normative choices.
& \pmark \\
& & d. Articulate relationship between proxies and real-world safety.
& \pmark \\

\cmidrule{2-4}

& \multirow{4}{*}{\textbf{R8. Locking and Versioning}}
& a. Record model access details and inference parameters.
& \pmark \\
& & b. Fix and report random seeds.
& \xmark \\
& & c. Version all evaluation components.
& \pmark \\
& & d. Provide documentation for independent replication.
& \cmark \\

\cmidrule{2-4}

& \multirow{4}{*}{\textbf{R9. Anchoring Proxies in Deployment}}
& a. State prompt sources and report validation if synthetic.
& \cmark \\
& & b. Justify why scenarios represent real-world use.
& \pmark \\
& & c. Include multi-turn evaluation for extended interaction risks.
& \xmark \\
& & d. Document trade-offs for in-the-wild collection constraints.
& \xmark \\

\cmidrule{2-4}

& \multirow{4}{*}{\textbf{R10. Iterative Refinement via Community}}
& a. Treat benchmarks as evolving instruments.
& \pmark \\
& & b. Involve affected communities in assessment.
& \xmark \\
& & c. State whether benchmark evaluates model alone or within system context.
& \pmark \\
& & d. Acknowledge model-level $\neq$ system-level safety.
& \pmark \\

\end{longtable}
\end{scriptsize}

\paragraph{Summary}
Of the 37 checklist items, AIR 2024 fully addresses 7 (\cmark), partially addresses 15 (\pmark), and does not address 15 (\xmark). The benchmark demonstrates strength in construct coverage documentation (R1) and value specification (R7), but shows gaps in dynamic risk discovery (R2), ML phenomena as safety concerns (R3), uncertainty quantification (R6), and community engagement (R10).

\section{Coding Details}
\label{app:coding}

\paragraph{Coding Dimensions}
We coded all surveyed benchmarks along multiple dimensions, including their assignment to Rumsfeld-style risk categories, whether benchmarks explicitly specify the risks not covered in their work, their reliance on fixed versus dynamic data, the use of binary outcome metrics, distinctions between harm severity levels and whether such distinctions are theoretically grounded, explicit treatment of uncertainty, grounding of proxies in external standards or regulations, and whether evaluations are limited to single-turn interactions. Coding protocol provided in additional file.

\newcommand{\colrotate}{45} 

\begin{scriptsize}
\begin{longtable}{p{2.0cm}c c c c c c c c c}
\caption{Coding results for all surveyed benchmarks.
One author performed initial coding; a second author reviewed all assignments, with disagreements resolved through discussion. 
\textbf{Column definitions}: 
\textbf{Rumsfeld}: Rumsfeld category assignment (KK = Known knowns, KU = Known unknowns, UK = Unknown knowns, UU = Unknown unknowns).
\textbf{Uncovered Doc.}: Whether the benchmark explicitly specifies the risks it uncovers, including unexpected, unknown, or unforeseen risks.
\textbf{Fixed Data}: Whether the benchmark relies on a fixed, static dataset.
\textbf{Binary Metric}: Whether the benchmark reduces evaluation to binary outcome proportions (e.g., harmful/safe, reject/not reject, biased/unbiased, attack success/failure) as the primary metric.
\textbf{Sev. Level}: Whether the benchmark distinguishes between different levels of harm severity (e.g., low/medium/high or Level~1--5).
\textbf{Sev. Grd.}: If severity levels are distinguished, whether they are grounded in an explicit theoretical framework or external standard (e.g., regulatory guidance or established harm taxonomies).
\textbf{Uncert.}: Whether the benchmark explicitly identifies sources of uncertainty or variation (e.g., model sensitivity, response stochasticity, sampling variation, or evaluator disagreement).
\textbf{Proxy Grd.}: Whether benchmark definitions or evaluation proxies are explicitly grounded in established frameworks, regulations, policies, or societal standards.
\textbf{Single-turn}: Whether the benchmark evaluates models in isolation via a single question-answer interaction, without contextual information or multiple trials.
\textbf{Notation.}
\cmark\ denotes yes,
\pmark\ denotes partial or mixed support,
\xmark\ denotes no,
and NA indicates not applicable.
}
\label{tab:full_coded_benchmarks} \\
\toprule
& \rotatebox{\colrotate}{\textbf{Rumsfeld}} 
& \rotatebox{\colrotate}{\textbf{Uncovered Doc.}} 
& \rotatebox{\colrotate}{\textbf{Fixed Data}} 
& \rotatebox{\colrotate}{\textbf{Binary Metric}} 
& \rotatebox{\colrotate}{\textbf{Sev. Level}} 
& \rotatebox{\colrotate}{\textbf{Sev. Grd.}} 
& \rotatebox{\colrotate}{\textbf{Uncert.}} 
& \rotatebox{\colrotate}{\textbf{Proxy Grd.}} 
& \rotatebox{\colrotate}{\textbf{Single-turn}} \\
\textbf{Benchmark} & & & & & & & & & \\
\midrule
\endfirsthead

\multicolumn{10}{c}{\scriptsize\tablename\ \thetable\ -- \textit{Continued from previous page}} \\
\toprule
& \rotatebox{\colrotate}{\textbf{Rumsfeld}} 
& \rotatebox{\colrotate}{\textbf{Uncovered Doc.}} 
& \rotatebox{\colrotate}{\textbf{Fixed Data}} 
& \rotatebox{\colrotate}{\textbf{Binary Metric}} 
& \rotatebox{\colrotate}{\textbf{Sev. Level}} 
& \rotatebox{\colrotate}{\textbf{Sev. Grd.}} 
& \rotatebox{\colrotate}{\textbf{Uncert.}} 
& \rotatebox{\colrotate}{\textbf{Proxy Grd.}} 
& \rotatebox{\colrotate}{\textbf{Single-turn}} \\
\textbf{Benchmark} & & & & & & & & & \\
\midrule
\endhead

\midrule
\multicolumn{10}{r}{\textit{Continued on next page}} \\
\endfoot

\bottomrule
\endlastfoot

CATQA~\citepbench{1CatQA} & KK & \pmark & \cmark & \cmark & \xmark & NA & \cmark & \cmark & \cmark \\
HOLISTICBIAS~\citepbench{2HolisticBias} & KK & \cmark & \cmark & \pmark & \xmark & NA & \cmark & \cmark & \cmark \\
SAFETEXT~\citepbench{3SafeText} & KK & \pmark & \cmark & \pmark & \xmark & NA & \cmark & \xmark & \cmark \\
WALLEDEVAL~\citepbench{4WalledEval} & KK & \pmark & \cmark & \cmark & \xmark & NA & \cmark & \pmark & \cmark \\
TOXIGEN~\citepbench{5Toxigen} & KK & \cmark & \cmark & \pmark & \cmark & \xmark & \cmark & \cmark & \cmark \\
JADE~\citepbench{6Jade} & KU & \pmark & \xmark & \pmark & \xmark & NA & \cmark & \pmark & \cmark \\
PROSOCIALDIALOG~\citepbench{7ProsocialDialog} & KK & \pmark & \cmark & \xmark & \cmark & \cmark & \cmark & \cmark & \xmark \\
StrongREJECT~\citepbench{8StrongREJECT} & KK & \cmark & \cmark & \xmark & \cmark & \xmark & \cmark & \cmark & \cmark \\
XSTEST~\citepbench{9XSTest} & KK & \pmark & \cmark & \cmark & \xmark & NA & \cmark & \xmark & \cmark \\
Cognitive Biases~\citepbench{10comprehensive} & KK & \pmark & \cmark & \xmark & \xmark & NA & \cmark & \cmark & \cmark \\
Non-Discrimination~\citepbench{11achieving} & KK & \xmark & \cmark & \xmark & \xmark & NA & \cmark & \xmark & \cmark \\
Societal Bias VLMs~\citepbench{13unified} & KK & \pmark & \cmark & \xmark & \xmark & NA & \cmark & \cmark & \cmark \\
AART~\citepbench{14AART} & KU & \cmark & \xmark & \cmark & \xmark & NA & \cmark & \cmark & \cmark \\
AAVENUE~\citepbench{15AAVENUE} & KK & \pmark & \cmark & \pmark & \xmark & NA & \cmark & \xmark & \cmark \\
Ego-View Accident~\citepbench{16abductive} & KK & \pmark & \cmark & \cmark & \xmark & NA & \cmark & \pmark & \cmark \\
Adversarial VQA~\citepbench{18AdversarialVQA} & KU & \xmark & \xmark & \cmark & \xmark & NA & \cmark & \xmark & \cmark \\
Adversarial GLUE~\citepbench{19AdversarialGLUE} & UK & \xmark & \cmark & \cmark & \xmark & NA & \cmark & \xmark & \cmark \\
SAFETY-TUNED LLAMAS~\citepbench{20Safety-TunedLLaMAs} & KK & \xmark & \cmark & \xmark & \cmark & \cmark & \cmark & \pmark & \cmark \\
AgentDojo~\citepbench{21AgentDojo} & KU & \cmark & \xmark & \cmark & \xmark & NA & \pmark & \xmark & \xmark \\
AILUMINATE~\citepbench{22AILuminate} & KK & \cmark & \cmark & \cmark & \xmark & NA & \cmark & \cmark & \cmark \\
AIR-BENCH 2024~\citepbench{23AIR-Bench2024} & KK & \pmark & \cmark & \pmark & \cmark & \cmark & \cmark & \cmark & \cmark \\
ALERT~\citepbench{24ALERT} & KK & \pmark & \cmark & \cmark & \xmark & NA & \cmark & \xmark & \cmark \\
QA-LIGN~\citepbench{25QA-LIGN} & KK & \cmark & \cmark & \pmark & \cmark & \xmark & \cmark & \xmark & \cmark \\
\citetbench{27wang2024all} & KK & \pmark & \cmark & \cmark & \xmark & NA & \cmark & \xmark & \cmark \\
\citetbench{29sotnikova2021analyzing} & KK & \cmark & \cmark & \pmark & \cmark & \xmark & \cmark & \cmark & \cmark \\
\citetbench{31alfadel2020threat} & KK & \cmark & \cmark & \cmark & \cmark & \cmark & \cmark & \cmark & NA \\
C2SaferRust~\citepbench{32C2SaferRust} & KK & \cmark & \cmark & \pmark & \xmark & NA & \cmark & \xmark & \xmark \\
\citetbench{33huang2022large} & KK & \cmark & \cmark & \cmark & \xmark & NA & \cmark & \cmark & \xmark \\
~\citepbench{35wan2023personalized} & KK & \xmark & \cmark & \pmark & \xmark & NA & \cmark & \xmark & \cmark \\
ArtPrompt~\citepbench{36ArtPrompt} & KU & \cmark & \xmark & \pmark & \cmark & \cmark & \pmark & \cmark & \cmark \\
WildTeaming~\citepbench{37WildTeaming} & KU & \cmark & \cmark & \cmark & \xmark & NA & \cmark & \xmark & \cmark \\
\citetbench{38huang2023catastrophic} & KU & \pmark & \cmark & \cmark & \xmark & NA & \cmark & \pmark & \cmark \\
Athena~\citepbench{39Athena} & KK & \cmark & \cmark & \cmark & \cmark & \xmark & \cmark & \xmark & \xmark \\
BackdoorLLM~\citepbench{40BackdoorLLM} & KU & \pmark & \cmark & \cmark & \xmark & NA & \cmark & \pmark & \xmark \\
BBG~\citepbench{41BBG} & KK & \pmark & \cmark & \cmark & \xmark & NA & \cmark & \xmark & \xmark \\
BBQ~\citepbench{42BBQ} & KK & \cmark & \cmark & \cmark & \xmark & NA & \cmark & \cmark & \cmark \\
BeaverTails~\citepbench{43BeaverTails} & KK & \xmark & \cmark & \pmark & \xmark & NA & \cmark & \xmark & \cmark \\
KOFFVQA~\citepbench{44KOFFVQA} & KK & \xmark & \cmark & \xmark & \cmark & \xmark & \cmark & \xmark & \cmark \\
\citetbench{45wang2025benchmark} & KU & \xmark & \xmark & \cmark & \xmark & NA & \cmark & \xmark & \cmark \\
Flames~\citepbench{46Flames} & KK & \xmark & \cmark & \pmark & \cmark & \cmark & \cmark & \cmark & \cmark \\
\citetbench{47_10376985} & KK & \xmark & \cmark & \cmark & \xmark & NA & \cmark & \xmark & \cmark \\
\citetbench{48verma2025hidden} & KK & \xmark & \cmark & \xmark & \xmark & NA & \cmark & \xmark & \cmark \\
\citetbench{49DBLP:conf/fat/LaszkiewiczDVFL24} & KK & \cmark & \cmark & \pmark & \xmark & NA & \cmark & \cmark & \cmark \\
SAFE~\citepbench{51SAFE} & KK & \xmark & \cmark & \pmark & \xmark & NA & \cmark & \xmark & \cmark \\
\citetbench{52kirk2021biasoutofthebox} & KK & \xmark & \cmark & \xmark & \xmark & NA & \cmark & \xmark & \cmark \\
BOLD~\citepbench{54BOLD} & KK & \cmark & \cmark & \pmark & \xmark & NA & \cmark & \xmark & \cmark \\
\citetbench{55xu-etal-2021-bot} & KK & \xmark & \cmark & \cmark & \cmark & \xmark & \cmark & \xmark & \xmark \\
\citetbench{56jaiswal2024breakingglobalnorthstereotype} & KK & \xmark & \cmark & \cmark & \xmark & NA & \cmark & \cmark & \cmark \\
\citetbench{57dinan-etal-2019-build} & KU & \xmark & \xmark & \cmark & \xmark & NA & \cmark & \xmark & \xmark \\
\citetbench{58-10.1145/3278721.3278767} & KK & \pmark & \xmark & \cmark & \xmark & NA & \cmark & \xmark & NA \\
CALM~\citepbench{59CALM} & KK & \cmark & \cmark & \cmark & \xmark & NA & \cmark & \cmark & \cmark \\
~\citetbench{60-10.1145/3627106.3627196} & KK & \pmark & \cmark & \pmark & \xmark & NA & \cmark & \pmark & \xmark \\
RuLES~\citepbench{61RuLES} & KK & \cmark & \cmark & \cmark & \xmark & NA & \cmark & \pmark & \xmark \\
CARNOVEL~\citepbench{62CARNOVEL} & UK & \pmark & \cmark & \pmark & \xmark & NA & \cmark & \xmark & \xmark \\
\citetbench{63-10.1145/3461702.3462614} & KK & \xmark & \cmark & \cmark & \xmark & NA & \cmark & \cmark & NA \\
OccuGender~\citepbench{64OccuGender} & KK & \xmark & \cmark & \xmark & \xmark & NA & \cmark & \xmark & \xmark \\
CDEval~\citepbench{65CDEval} & KK & \xmark & \cmark & \pmark & \xmark & NA & \cmark & \cmark & \cmark \\
CHBias~\citepbench{66CHBias} & KK & \xmark & \cmark & \xmark & \xmark & NA & \cmark & \pmark & \xmark \\
CHiSafetyBench~\citepbench{67CHiSafetyBench} & KK & \xmark & \cmark & \pmark & \cmark & \cmark & \cmark & \cmark & \xmark \\
CIF-Bench~\citepbench{68CIF-Bench} & KK & \xmark & \cmark & \pmark & \xmark & NA & \cmark & \xmark & \cmark \\
CBBQ~\citepbench{69CBBQ} & KK & \xmark & \cmark & \cmark & \cmark & NA & \cmark & \cmark & \cmark \\
OW-DFA~\citepbench{70OW-DFA} & KU & \cmark & \xmark & \cmark & \xmark & NA & \cmark & \xmark & \cmark \\
CPO~\citepbench{71CPO} & KK & \pmark & \cmark & \xmark & \cmark & \xmark & \cmark & \xmark & \xmark \\
CoSafe~\citepbench{72CoSafe} & KK & \xmark & \cmark & \cmark & \xmark & NA & \cmark & \xmark & \xmark \\
CrowS-Pairs~\citepbench{73CrowS-Pairs} & KK & \pmark & \cmark & \cmark & \xmark & NA & \cmark & \cmark & \cmark \\
CVE-Bench~\citepbench{74CVE-Bench} & KK & \xmark & \cmark & \cmark & \xmark & NA & \cmark & \cmark & \xmark \\
Cybench~\citepbench{75Cybench} & KK & \pmark & \cmark & \pmark & \xmark & NA & \pmark & \cmark & \xmark \\
\citetbench{76-10.1145/3442188.3445896} & KK & \pmark & \xmark & \xmark & \xmark & NA & \cmark & \cmark & \xmark \\
DELPHI~\citepbench{77DELPHI} & KK & \xmark & \cmark & \pmark & \xmark & NA & \cmark & \xmark & \cmark \\
DICES~\citepbench{79DICES} & KK & \pmark & \cmark & \pmark & \cmark & \xmark & \cmark & \xmark & \xmark \\
~\citetbench{80-perez-etal-2023-discovering} & UU & \cmark & \cmark & \pmark & \xmark & NA & \cmark & \cmark & \cmark \\
discrim-eval~\citepbench{81discrim-eval} & KK & \pmark & \cmark & \pmark & \xmark & NA & \cmark & \cmark & \cmark \\
~\citetbench{82-10350388} & KK & \pmark & \cmark & \xmark & \xmark & NA & \cmark & \xmark & \cmark \\
DiversityMedQA~\citepbench{83DiversityMedQA} & KK & \xmark & \cmark & \cmark & \xmark & NA & \cmark & \cmark & \cmark \\
JailbreakHub~\citepbench{84JailbreakHub} & KU & \pmark & \cmark & \cmark & \cmark & \cmark & \cmark & \cmark & \xmark \\
Do-Not-Answer~\citepbench{85Do-Not-Answer} & KK & \cmark & \cmark & \pmark & \cmark & \xmark & \cmark & \xmark & \cmark \\
MACHIAVELLI~\citepbench{86MACHIAVELLI} & KU & \cmark & \xmark & \cmark & \cmark & \cmark & \cmark & \xmark & \xmark \\
~\citetbench{87-liu-etal-2020-gender} & KK & \xmark & \cmark & \pmark & \xmark & NA & \cmark & \xmark & \cmark \\
ConfAIde~\citepbench{88ConfAIde} & KK & \cmark & \cmark & \pmark & \xmark & NA & \xmark & \cmark & \xmark \\
LabellessFace~\citepbench{89LabellessFace} & KK & \cmark & \cmark & \cmark & \xmark & NA & \cmark & \xmark & \cmark \\
ePiC~\citepbench{90ePiC} & KK & \cmark & \cmark & \cmark & \xmark & NA & \cmark & \xmark & \cmark \\
GenMO~\citepbench{92GenMO} & KK & \xmark & \cmark & \cmark & \xmark & NA & \cmark & \xmark & \cmark \\
RealToxicityPrompts~\citepbench{93RealToxicityPrompts} & KK & \cmark & \cmark & \pmark & \xmark & NA & \cmark & \xmark & \cmark \\
Tri-HE~\citepbench{94Tri-HE} & KK & \xmark & \cmark & \cmark & \xmark & NA & \cmark & \xmark & \cmark \\
~\citetbench{95b-li-etal-2024-evaluating-instruction} & KU & \pmark & \cmark & \pmark & \xmark & NA & \cmark & \xmark & \cmark \\
~\citetbench{96-kiritchenko-mohammad-2018-examining} & KK & \cmark & \cmark & \xmark & \xmark & NA & \cmark & \pmark & \cmark \\
AdvMark~\citepbench{97AdvMark} & KU & \xmark & \cmark & \pmark & \xmark & NA & \cmark & \cmark & \cmark \\
FACET~\citepbench{98FACET} & KK & \pmark & \cmark & \xmark & \xmark & NA & \cmark & \cmark & \cmark \\
FairLex~\citepbench{99FairLex} & KK & \pmark & \cmark & \xmark & \xmark & NA & \cmark & \cmark & \cmark \\
~\citetbench{100-siddiqui2022biasfacialanalysis} & KK & \xmark & \cmark & \pmark & \xmark & NA & \xmark & \cmark & \cmark \\
FFT~\citepbench{101FFT} & KK & \xmark & \cmark & \pmark & \xmark & NA & \cmark & \pmark & \cmark \\
Filipino~\citepbench{102Filipino} & KK & \pmark & \cmark & \cmark & \xmark & NA & \cmark & \pmark & \cmark \\
FLEX~\citepbench{103FLEX} & KU & \pmark & \cmark & \cmark & \xmark & NA & \cmark & \xmark & \cmark \\
FLAIR~\citepbench{104FLAIR} & KK & \cmark & \cmark & \xmark & \xmark & NA & \cmark & \xmark & \cmark \\
FrenchCrowS-Pairs~\citepbench{105FrenchCrowS-Pairs} & KK & \pmark & \cmark & \cmark & \xmark & NA & \cmark & \pmark & \cmark \\
FrenchToxicityPrompts~\citepbench{106FrenchToxicityPrompts} & KK & \cmark & \cmark & \pmark & \cmark & \xmark & \cmark & \xmark & \cmark \\
Z'eroe~\citepbench{107Z'eroe} & KK & \pmark & \cmark & \pmark & \xmark & NA & \cmark & \xmark & \cmark \\
WinoBias~\citepbench{108WinoBias} & KK & \xmark & \cmark & \cmark & \xmark & NA & \cmark & \cmark & \cmark \\
GeoNet~\citepbench{109GeoNet} & KK & \cmark & \cmark & \cmark & \xmark & NA & \cmark & \xmark & \cmark \\
CPAD~\citepbench{110CPAD} & KK & \xmark & \cmark & \cmark & \xmark & NA & \cmark & \pmark & \xmark \\
GPTFuzz~\citepbench{111GPTFuzz} & KU & \pmark & \xmark & \cmark & \xmark & NA & \cmark & \xmark & \cmark \\
CoP~\citepbench{112CoP} & KU & \cmark & \xmark & \cmark & \cmark & \xmark & \cmark & \pmark & \xmark \\
SafeWatch-Bench~\citepbench{113SafeWatch-Bench} & KK & \cmark & \cmark & \pmark & \xmark & NA & \cmark & \cmark & \xmark \\
HarmBench~\citepbench{114HarmBench} & KU & \cmark & \cmark & \cmark & \xmark & NA & \cmark & \cmark & \xmark \\
HRS-Bench~\citepbench{115HRS-Bench} & KK & \xmark & \cmark & \xmark & \xmark & NA & \cmark & \pmark & \cmark \\
FVQA2.0~\citepbench{116FVQA2.0} & KU & \pmark & \cmark & \cmark & \xmark & NA & \cmark & \pmark & \cmark \\
HypoTermQA~\citepbench{117HypoTermQA} & KU & \pmark & \cmark & \cmark & \xmark & NA & \cmark & \xmark & \cmark \\
IHEval~\citepbench{118IHEval} & KK & \cmark & \cmark & \pmark & \xmark & NA & \pmark & \cmark & \xmark \\
IndiBias~\citepbench{119IndiBias} & KK & \pmark & \cmark & \pmark & \xmark & NA & \cmark & \pmark & \cmark \\
UNQOVER~\citepbench{120UNQOVER} & KK & \pmark & \cmark & \xmark & \xmark & NA & \cmark & \pmark & \cmark \\
InjecAgent~\citepbench{121InjecAgent} & KK & \pmark & \cmark & \cmark & \xmark & NA & \cmark & \xmark & \xmark \\
ADVQA~\citepbench{122ADVQA} & KU & \pmark & \cmark & \xmark & \xmark & NA & \cmark & \cmark & \cmark \\
JailbreakBench~\citepbench{123JailbreakBench} & KK & \pmark & \xmark & \cmark & \xmark & NA & \cmark & \cmark & \xmark \\
~\citetbench{124-ghanim-etal-2024-jailbreaking} & KK & \cmark & \cmark & \cmark & \xmark & NA & \cmark & \xmark & \xmark \\
JailBreakV-28K~\citepbench{125JailBreakV-28K} & KK & \xmark & \cmark & \cmark & \xmark & NA & \xmark & \cmark & \cmark \\
JobFair~\citepbench{126JobFair} & KK & \cmark & \cmark & \pmark & \cmark & \xmark & \cmark & \cmark & \xmark \\
KoBBQ~\citepbench{127KoBBQ} & KK & \pmark & \cmark & \cmark & \xmark & NA & \cmark & \cmark & \cmark \\
KorNAT~\citepbench{128KorNAT} & KK & \pmark & \cmark & \pmark & \xmark & NA & \cmark & \cmark & \cmark \\
LatentJailbreak~\citepbench{129LatentJailbreak} & KK & \pmark & \cmark & \cmark & \xmark & NA & \cmark & \xmark & \xmark \\
CCLR~\citepbench{130CCLR} & UK & \xmark & \cmark & \xmark & \xmark & NA & \xmark & \xmark & NA \\
~\citetbench{131a-jain2025interactinglargelanguagemodel} & KK & \pmark & \xmark & \xmark & \cmark & \cmark & \cmark & \pmark & \xmark \\
LLMArena~\citepbench{131bLLMArena} & UU & \pmark & \xmark & \xmark & \xmark & NA & \cmark & \xmark & \xmark \\
SAD~\citepbench{133SAD} & KK & \cmark & \cmark & \cmark & \xmark & NA & \cmark & \pmark & \cmark \\
MedHALT~\citepbench{135MedHALT} & KK & \pmark & \cmark & \cmark & \xmark & NA & \cmark & \cmark & \cmark \\
MEDFAIR~\citepbench{136MEDFAIR} & KK & \pmark & \cmark & \xmark & \xmark & NA & \cmark & \xmark & \cmark \\
MedSafetyBench~\citepbench{137MedSafetyBench} & KK & \pmark & \cmark & \xmark & \cmark & \cmark & \cmark & \cmark & \cmark \\
~\citetbench{139-ahn-oh-2021-mitigating} & KK & \pmark & \cmark & \xmark & \xmark & NA & \cmark & \pmark & \cmark \\
MLeVLM~\citepbench{140MLeVLM} & KK & \pmark & \cmark & \pmark & \cmark & NA & \cmark & \pmark & \xmark \\
MMEvalPro~\citepbench{141MMEvalPro} & KK & \pmark & \cmark & \cmark & \xmark & NA & \cmark & \xmark & \xmark \\
ModSCAN~\citepbench{142ModSCAN} & KK & \cmark & \cmark & \pmark & \xmark & NA & \cmark & \cmark & \cmark \\
MultiRobustBench~\citepbench{143MultiRobustBench} & KK & \cmark & \cmark & \pmark & \xmark & NA & \cmark & \xmark & \cmark \\
~\citetbench{145-rudinger-etal-2018-gender} & KK & \pmark & \cmark & \cmark & \xmark & NA & \cmark & \cmark & \cmark \\
OKTest~\citepbench{146OKTest} & KK & \cmark & \cmark & \cmark & \xmark & NA & \cmark & \xmark & \cmark \\
CLCA~\citepbench{147CLCA} & KK & \cmark & \cmark & \pmark & \xmark & NA & \cmark & \cmark & \xmark \\
~\citetbench{148-shaikh-etal-2023-second} & KK & \xmark & \cmark & \cmark & \xmark & NA & \cmark & \xmark & \xmark \\
~\citetbench{149-sun-etal-2022-safety} & KK & \cmark & \cmark & \cmark & \xmark & NA & \cmark & \cmark & \xmark \\
~\citetbench{150-vashishtha-etal-2023-evaluating} & KK & \pmark & \cmark & \cmark & \xmark & NA & \cmark & \xmark & \cmark \\
CAMeL-2~\citepbench{151CAMeL-2} & KK & \cmark & \cmark & \cmark & \xmark & NA & \cmark & \pmark & \cmark \\
~\citetbench{152-DBLP:conf/cvpr/ChatterjeeGBY24} & KK & \pmark & \cmark & \xmark & \xmark & NA & \cmark & \xmark & \cmark \\
OR-Bench~\citepbench{153OR-Bench} & KU & \pmark & \xmark & \cmark & \xmark & NA & \cmark & \cmark & \cmark \\
PAPILLON~\citepbench{154PAPILLON} & KK & \cmark & \cmark & \cmark & \xmark & NA & \cmark & \xmark & \cmark \\
PERSONA~\citepbench{156PERSONA} & KK & \cmark & \cmark & \cmark & \xmark & NA & \cmark & \cmark & \cmark \\
PLUE~\citepbench{157PLUE} & KK & \pmark & \cmark & \cmark & \xmark & NA & \cmark & \pmark & \cmark \\
Poser~\citepbench{158Poser} & KK & \pmark & \cmark & \cmark & \xmark & NA & \cmark & \cmark & \cmark \\
PrivLM-Bench~\citepbench{159PrivLM-Bench} & KK & \xmark & \cmark & \pmark & \xmark & NA & \cmark & \cmark & \cmark \\
PROMPTEVALS~\citepbench{160PROMPTEVALS} & KK & \pmark & \cmark & \xmark & \xmark & NA & \cmark & \cmark & \cmark \\
MLLMU-Bench~\citepbench{161MLLMU-Bench} & KK & \cmark & \cmark & \pmark & \cmark & NA & \xmark & \cmark & \cmark \\
RADDLE~\citepbench{162RADDLE} & KU & \xmark & \cmark & \cmark & \xmark & NA & \cmark & \pmark & \xmark \\
RAGTruth~\citepbench{163RAGTruth} & KK & \pmark & \cmark & \cmark & \cmark & NA & \cmark & \xmark & \xmark \\
~\citetbench{164-bhatt-etal-2022-contextualizing} & KK & \cmark & \cmark & \xmark & \xmark & NA & \cmark & \pmark & \cmark \\
MDIT-Bench~\citepbench{165MDIT-Bench} & KK & \cmark & \cmark & \cmark & \xmark & NA & \cmark & \cmark & \xmark \\
REAP~\citepbench{166REAP} & KK & \xmark & \cmark & \pmark & \xmark & NA & \cmark & \cmark & \cmark \\
~\citetbench{167-ganguli2022redteaminglanguagemodels} & KK & \cmark & NA & \pmark & \cmark & NA & \cmark & \pmark & \xmark \\
RED-EVAL~\citepbench{168RED-EVAL} & KU & \pmark & \cmark & \cmark & \xmark & NA & \cmark & \pmark & \xmark \\
RedditBias~\citepbench{169RedditBias} & KK & \pmark & \cmark & \xmark & \xmark & NA & \cmark & \cmark & \xmark \\
DeMET~\citepbench{170DeMET} & KK & \xmark & \cmark & \pmark & \xmark & NA & \cmark & \cmark & \cmark \\
TLDR~\citepbench{171TLDR} & KK & \xmark & \cmark & \pmark & \xmark & NA & \cmark & \pmark & \cmark \\
ROBBIE~\citepbench{172ROBBIE} & KK & \cmark & \cmark & \cmark & \xmark & NA & \cmark & \pmark & \cmark \\
Robo3D~\citepbench{173Robo3D} & KK & \pmark & \cmark & \xmark & \cmark & \cmark & \cmark & \pmark & \cmark \\
GQA-OOD~\citepbench{174GQA-OOD} & KK & \pmark & \cmark & \cmark & \xmark & NA & \cmark & \xmark & \cmark \\
S-Eval~\citepbench{175S-Eval} & KK & \pmark & \xmark & \cmark & \xmark & NA & \cmark & \cmark & \xmark \\
LaMaSafe~\citepbench{176LaMaSafe} & KK & \pmark & \cmark & \xmark & \xmark & NA & \cmark & \xmark & \xmark \\
SafeBench~\citepbench{177SafeBench} & KU & \xmark & \xmark & \pmark & \xmark & NA & \pmark & \cmark & \xmark \\
~\citetbench{178-journals/corr/abs-2304-10436} & KK & \pmark & \cmark & \cmark & \xmark & NA & \cmark & \xmark & \cmark \\
SafetyBench~\citepbench{179SafetyBench} & KK & \cmark & \cmark & \cmark & \xmark & NA & \cmark & \xmark & \cmark \\
SALAD-Bench~\citepbench{180SALAD-Bench} & KK & \pmark & \cmark & \cmark & \xmark & NA & \cmark & \cmark & \xmark \\
~\citetbench{181-sun-miceli-barone-2024-scaling} & KK & \cmark & \cmark & \pmark & \xmark & NA & \cmark & \xmark & \cmark \\
SG-Bench~\citepbench{182SG-Bench} & KK & \pmark & \cmark & \cmark & \xmark & NA & \cmark & \pmark & \xmark \\
SimpleSafetyTests~\citepbench{183SimpleSafetyTests} & KK & \cmark & \cmark & \cmark & \xmark & NA & \cmark & \cmark & \cmark \\
SoFa~\citepbench{184SoFa} & KK & \cmark & \cmark & \xmark & \xmark & NA & \cmark & \pmark & \cmark \\
~\citetbench{185-conf/coling/SadhuSS25} & KK & \pmark & \cmark & \cmark & \xmark & NA & \cmark & \pmark & \cmark \\
SORRY-Bench~\citepbench{186SORRY-Bench} & KK & \cmark & \cmark & \cmark & \xmark & NA & \cmark & \cmark & \cmark \\
StereoSet~\citepbench{187StereoSet} & KK & \cmark & \cmark & \cmark & \xmark & NA & \cmark & \xmark & \xmark \\
BenchPress~\citepbench{188BenchPress} & KK & \cmark & \cmark & \xmark & \xmark & NA & \cmark & \xmark & \xmark \\
TensorTrust~\citepbench{189TensorTrust} & KU & \cmark & \cmark & \cmark & \xmark & NA & \cmark & \xmark & \cmark \\
COCONOT~\citepbench{191COCONOT} & KK & \pmark & \cmark & \cmark & \xmark & NA & \cmark & \cmark & \cmark \\
~\citetbench{192-conf/emnlp/AakankshaAEGKFH24} & KK & \cmark & \cmark & \cmark & \xmark & NA & \cmark & \xmark & \cmark \\
SwedishWinogender~\citepbench{193SwedishWinogender} & KK & \pmark & \cmark & \cmark & \xmark & NA & \cmark & \cmark & \cmark \\
WMDP~\citepbench{194WMDP} & KK & \pmark & \cmark & \cmark & \xmark & NA & \cmark & \cmark & \cmark \\
~\citetbench{195-conf/emnlp/ShengCNP19} & KK & \xmark & \cmark & \xmark & \xmark & NA & \cmark & \pmark & \cmark \\
~\citetbench{197-journals/corr/abs-1710-06881} & UK & \cmark & NA & \xmark & \xmark & NA & \xmark & \xmark & NA \\
TheGreatestGood~\citepbench{198TheGreatestGood} & KK & \pmark & \cmark & \xmark & \xmark & NA & \cmark & \cmark & \cmark \\
TAB~\citepbench{199TAB} & KU & \pmark & \cmark & \pmark & \cmark & \cmark & \cmark & \cmark & NA \\
EmpatheticDialogues~\citepbench{200EmpatheticDialogues} & KK & \xmark & \cmark & \xmark & \cmark & \xmark & \cmark & \cmark & \xmark \\
~\citetbench{201-conf/emnlp/ZhouDMLWHJ0M22} & KK & \cmark & \cmark & \cmark & \xmark & NA & \cmark & \xmark & \xmark \\
MMHB~\citepbench{202MMHB} & KK & \pmark & \cmark & \pmark & \xmark & NA & \xmark & \pmark & \cmark \\
~\citetbench{203-conf/icml/LiangWMS21} & KK & \pmark & \cmark & \xmark & \xmark & NA & \cmark & \xmark & \xmark \\
~\citetbench{204-conf/cvpr/LiZCWFWL0023} & KK & \xmark & \cmark & \xmark & \xmark & NA & \cmark & \xmark & \cmark \\
~\citetbench{205-conf/coling/SantoshBSGN24} & KK & \pmark & \cmark & \pmark & \xmark & NA & \cmark & \xmark & \cmark \\
~\citetbench{206-bai2022helpfulharmlessassistant} & KK & \cmark & \xmark & \cmark & \xmark & NA & \cmark & \xmark & \xmark \\
TrustLLM~\citepbench{207TrustLLM} & KK & \xmark & \cmark & \pmark & \cmark & \cmark & \cmark & \cmark & \xmark \\
TruthfulQA~\citepbench{208TruthfulQA} & KK & \pmark & \cmark & \pmark & \cmark & \xmark & \cmark & \pmark & \cmark \\
~\citetbench{209-conf/iclr/Qi0XC0M024} & KU & \cmark & \cmark & \pmark & \cmark & \xmark & \cmark & \cmark & \cmark \\
CII-Bench~\citepbench{210CII-Bench} & KK & \pmark & \cmark & \pmark & \cmark & \xmark & \cmark & \xmark & \cmark \\
UNQOVER~\citepbench{211UNQOVER} & KK & \cmark & \cmark & \pmark & \xmark & NA & \cmark & \xmark & \cmark \\
~\citetbench{212-conf/iccv/ZhaoWR21} & KK & \cmark & \cmark & \xmark & \xmark & NA & \cmark & \cmark & \cmark \\
MHaluBench~\citepbench{213MHaluBench} & KK & \cmark & \cmark & \cmark & \xmark & NA & \cmark & \xmark & \cmark \\
AdvBench~\citepbench{214AdvBench} & KU & \pmark & \cmark & \pmark & \xmark & NA & \cmark & \xmark & \cmark \\
~\citetbench{215-journals/corr/abs-2311-04124} & KU & \pmark & \cmark & \xmark & \xmark & NA & \cmark & \pmark & \cmark \\
LMMs-Eval~\citepbench{216LMMs-Eval} & UK & \pmark & \xmark & \xmark & \xmark & NA & \cmark & \cmark & \xmark \\
VALUE~\citepbench{217VALUE} & KK & \cmark & \cmark & \pmark & \xmark & NA & \cmark & \xmark & \cmark \\
MixCuBe~\citepbench{218MixCuBe} & KK & \xmark & \cmark & \cmark & \xmark & NA & \cmark & \xmark & \cmark \\
B-score~\citepbench{219B-score} & KK & \xmark & \cmark & \pmark & \xmark & NA & \cmark & \xmark & \xmark \\
WinoQueer~\citepbench{220WinoQueer} & KK & \cmark & \cmark & \cmark & \xmark & NA & \cmark & \cmark & \cmark \\
SeqAR~\citepbench{221SeqAR} & KU & \pmark & \cmark & \cmark & \xmark & NA & \cmark & \xmark & \cmark \\
~\citetbench{222-journals/corr/abs-2409-00551} & KU & \pmark & \xmark & \cmark & \xmark & NA & \cmark & \xmark & \cmark \\
GEST~\citepbench{223GEST} & KK & \cmark & \cmark & \pmark & \xmark & NA & \cmark & \cmark & \cmark \\
TiEBe~\citepbench{224TiEBe} & KK & \cmark & \cmark & \cmark & \xmark & NA & \cmark & \xmark & \cmark \\
WorldCuisines~\citepbench{225WorldCuisines} & KK & \pmark & \cmark & \pmark & \xmark & NA & \cmark & \xmark & \cmark \\
VITAL~\citepbench{226VITAL} & KK & \cmark & \cmark & \pmark & \xmark & NA & \cmark & \pmark & \cmark \\
\end{longtable}
\end{scriptsize}

{\footnotesize
\bibliographystylebench{unsrtnat}
\bibliographybench{benchmark_references}
}





\end{document}